\documentclass[ApJ]{emulateapj}
\usepackage{graphicx,dcolumn,bm,epsfig,graphicx,hyperref,amsmath,amssymb}
\usepackage{comment}
\usepackage{placeins}
\usepackage{url}
\usepackage{ulem}
\newcommand{\dtm}{\textit{DTM}}

\usepackage[usenames,dvipsnames]{color}

\usepackage{url}
\usepackage{graphicx}

\begin{document}


\shortauthors{Sparre et al.}

\shorttitle{Dust and Metals in a GRB host galaxy at redshift 5}
\title{The metallicity and dust content of a redshift 5 gamma-ray burst host galaxy}



\author{
M.~Sparre\altaffilmark{1}, 
O.~E.~Hartoog\altaffilmark{2}, 
T.~Kr\"uhler\altaffilmark{1,3}, 
J.~P.~U. Fynbo\altaffilmark{1},
D.~J. Watson\altaffilmark{1},
K.~Wiersema\altaffilmark{4},
V.~D'Elia\altaffilmark{5,6}, 
T. Zafar\altaffilmark{7},
P.~M. J. Afonso\altaffilmark{8},
S.~Covino\altaffilmark{9},
A.~de Ugarte Postigo\altaffilmark{1,10},
H.~Flores\altaffilmark{11},
P.~Goldoni\altaffilmark{12}, 
J.~Greiner\altaffilmark{13},
J.~Hjorth\altaffilmark{1},
P.~Jakobsson\altaffilmark{14}, 
L.~Kaper\altaffilmark{2},
S.~Klose\altaffilmark{15},
A.~J.~Levan\altaffilmark{16},
D.~Malesani\altaffilmark{1},
B.~Milvang-Jensen\altaffilmark{1}, 
M.~Nardini\altaffilmark{17},
S.~Piranomonte\altaffilmark{18},
J.~Sollerman\altaffilmark{19},
R.~S\'anchez-Ram\'irez\altaffilmark{10},
S.~Schulze\altaffilmark{20},
N.~R. Tanvir\altaffilmark{4},
S.~D.~Vergani\altaffilmark{9,11},\\
R.~A.~M.~J. Wijers\altaffilmark{2}
}

\email{sparre@dark-cosmology.dk}

\altaffiltext{1}{Dark Cosmology Centre, Niels Bohr Institute, University of Copenhagen, Juliane Maries Vej 30, 2100 Copenhagen, Denmark}
\altaffiltext{2}{Anton Pannekoek Institute for Astronomy, University of Amsterdam, Science Park 904, 1098 XH Amsterdam}
\altaffiltext{3}{European Southern Observatory, Alonso de C\'{o}rdova 3107, Vitacura, Casilla 19001, Santiago 19, Chile}
\altaffiltext{4}{Department of Physics and Astronomy, University of Leicester, University Road, Leicester, LE1 7RH, UK}
\altaffiltext{5}{INAF / Rome Astronomical Observatory, via Frascati 33, 00040, Monteporzio Catone (Roma), Italy}
\altaffiltext{6}{ASI Science Data Centre, Via Galileo Galilei, 00044, Frascati (Roma) Italy}
\altaffiltext{7}{European Southern Observatory, Karl-Schwarzschild-Strasse 2, 85748, Garching, Germany}
\altaffiltext{8}{American River College, Physics and Astronomy Dpt., 4700 College Oak Drive, Sacramento, CA 95841, USA}
\altaffiltext{9}{INAF, Osservatorio Astronomico di Brera, via E. Bianchi 46, 23807 Merate (LC), Italy}
\altaffiltext{10}{Instituto de Astrof\'isica de Andaluc\'ia, CSIC, Glorieta de la Astronom\'ia s/n, E - 18008 Granada, Spain}
\altaffiltext{11}{Laboratoire GEPI, Observatoire de Paris, CNRS-UMR8111, Univ. Paris-Diderot 5 place Jules Janssen, 92195 Meudon France}
\altaffiltext{12}{APC, Astroparticule et Cosmologie, Universite Paris Diderot, CNRS/IN2P3, CEA/Irfu, Observatoire de Paris, Sorbonne Paris Cité, 10, Rue Alice Domon et L\'eonie Duquet, 75205, Paris Cedex 13, France}
\altaffiltext{13}{Max-Planck-Institut f\"{u}r extraterrestrische Physik, Giessenbachstra\ss e, 85748 Garching, Germany}
\altaffiltext{14}{Centre for Astrophysics and Cosmology, Science Institute, University of Iceland, Dunhagi 5, IS-107 Reykjavik, Iceland}
\altaffiltext{15}{Th\"uringer Landessternwarte Tautenburg, Sternwarte 5, 07778 Tautenburg, Germany}
\altaffiltext{16}{Department of Physics, University of Warwick, Coventry CV4 7AL, UK}
\altaffiltext{17}{Universit\`a degli studi di Milano-Bicocca, Piazza della Scienza 3, 20126, Milano, Italy}
\altaffiltext{18}{INAF-Osservatorio Astronomico di Roma, Via Frascati 33, I-00040 Monte Porzio Catone, Roma, Italy}
\altaffiltext{19}{Oskar Klein Centre, Department of Astronomy, AlbaNova, Stockholm University, 106 91 Stockholm, Sweden}
\altaffiltext{20}{Instituto de Astrof\'{i}sica, Facultad de F\'{i}sica, Pontificia Universidad Cat\'{o}lica de Chile, Casilla 306, Santiago 22, Chile Millennium Center for Supernova Science}


\begin{abstract}
Observations of the afterglows of long gamma-ray bursts (GRBs) allow
the study of star-forming galaxies across most of cosmic history. Here
we present observations of GRB\,111008A from which we can measure
metallicity, chemical abundance patterns, dust-to-metals ratio and extinction of the GRB host
galaxy at $z=5.0$. The host absorption system is a damped Lyman-$\alpha$ 
absorber (DLA) with a very large neutral hydrogen column density of 
$\log N(\mbox{\ion{H}{1}})/\text{cm}^{-2}=22.30\pm0.06$, and a metallicity of $[$S$/$H$]= -1.70 \pm 0.10$. It is the highest redshift GRB with such a precise metallicity measurement. The presence of fine-structure lines confirms the $z=5.0$ system as the GRB host galaxy, and makes this the highest redshift where \ion{Fe}{2} fine-structure lines have been detected. The afterglow is mildly reddened with $A_V = 0.11 \pm 0.04$ mag, and the host galaxy has a dust-to-metals ratio which is consistent with being equal to or lower than typical values in the Local Group.
\end{abstract}


\section{Introduction}

The study of gamma-ray bursts (GRBs) has an impact on a wide range of topics in
astrophysics. It is now well established that long-duration
GRBs originate from the collapse of massive stars
\citep[e.g.,][]{1998Natur.395..670G,2003Natur.423..847H,2003ApJ...591L..17S,2006ApJ...645L..21M,2006Natur.442.1008C,2011ApJ...735L..24S,2011ApJ...743..204B,2012grbu.book..169H,2013arXiv1305.6832X},
although we have yet to elucidate the detailed nature of the progenitor systems. 

Spectroscopy of GRB afterglows requires the most advanced and largest ground-based optical to near-infrared (NIR) 
telescopes, especially when these events occur at high redshift, since the source is weaker and the rest-frame UV absorption lines we detect are shifted to longer wavelengths. GRB afterglows can probe star forming regions out to redshifts of $z=8$--$9$ \citep{2009Natur.461.1254T,2009Natur.461.1258S,2011ApJ...736....7C}, and allow us to study galaxies in the early Universe that would normally have been too faint to be detected \citep{2012ApJ...754...46T,2012A&A...542A.103B}. In some cases it is possible to
determine column densities of both \ion{H}{1} and metals in
their host galaxies and hence calculate abundances for a wide
range of chemical elements
\citep[e.g.,][]{2006A&A...451L..47F,2006NJPh....8..195S,2007ApJ...666..267P,2009A&A...506..661L,2010A&A...523A..36D,2013MNRAS.428.3590T}.

Observational selection effects restrict the samples of GRBs for which chemical enrichment can be probed by this means. For many GRBs, even when spectroscopy is obtained, it proves insufficient to probe the rest-frame UV due to unfortunate redshifts that leave the relevant transitions (in particular
\ion{H}{1}) out of the observable range. In other cases, dust extinction within the host galaxies (the likely explanation for many `dark' bursts, see \citealt{2004ApJ...617L..21J,2009AJ....138.1690P,2011A&A...534A.108K}) makes the afterglows too faint for useful spectroscopy. Therefore the sample of bursts with afterglow spectroscopy is
not representative for all GRBs
\citep[e.g.,][]{2009ApJS..185..526F,2013A&A...557A..18K}.

With observations of GRB afterglows one can hope to
obtain the imprints in the ISM of chemical enrichment from core-collapse
supernovae (SNe), which are believed to dominate the metal production for very
young systems \citep[e.g.,][]{1986A&A...154..279M}.
The signature of such a chemical enrichment profile is an over-abundance of 
$\alpha$-elements as seen in metal-poor stars in the Local Group and in $
z>4$ DLAs\footnote{DLA: damped Lyman-$\alpha$ absorber: a sight line absorber with $\log N(\mbox{\ion{H}{1}})/\text{cm}^{-2}>20.30$ \citep{2005ARA&A..43..861W}}. \citep[e.g.,][]{2011Sci...333..176T,2012ApJ...755...89R}.

The aim of this article is to study the metal and dust properties of the host galaxy of GRB\,111008A at $z=5.0$, which provides a rare opportunity of measuring such characteristics for a high-redshift star-forming galaxy. The paper is structured as follows. In Section~\ref{sec_data} we present our observations, in Section~\ref{sec_metallicity}
we determine the metallicity of the host galaxy, and we also
discuss the presence and strength of fine-structure lines. In Section~\ref{sec_grond} we describe the spectral energy distribution (SED) of the NIR-to-X-ray data. In
Section~\ref{sec_intervening} we investigate an intervening DLA at $z=4.6$, in Section \ref{EmissionSearch} we present an attempt to identify the intervening absorber and the GRB host galaxy in emission and in Section~\ref{DustAndMetals} we study the amount of dust and metals in the host galaxy. 

For the cosmological calculations we assume a $\Lambda$CDM-universe with
$\Omega_\Lambda = 0.73$, $\Omega_\mathrm{m} = 0.27$, and $h_0=0.71$. We use $1\sigma$ error bars and $2\sigma$ upper and lower limits.

\section{Observations}\label{sec_data}

GRB\,111008A was discovered by \textit{Swift}/BAT \citep{2011GCN..12423...1S} with a duration of $T_{90}=63.46\pm 2.19$ s \citep{2011GCN..12424...1B}. An X-ray counterpart was subsequently discovered with \textit{Swift}/XRT \citep{2011GCN..12425...1B}, and the optical counterpart was later observed using several instruments \citep{2011GCN..12426...1L,2011GCN..12427...1X,2011GCN..12428...1N} before a spectroscopic redshift of $z=5.0$ was identified with Gemini/GMOS \citep{2011GCN..12429...1L}. This was later confirmed with VLT/X-shooter, and an intervening absorber at $z=4.6$ was also identified \citep{2011GCN..12431...1W}. The details of our observations are given below.

\subsection{X-shooter spectroscopy}

We obtained optical/NIR spectroscopy of the afterglow
of GRB~111008A with X-shooter \citep{2011A&A...536A.105V} mounted at UT2 of the
European Southern Observatory's (ESO) Very Large Telescope. Our spectrum
covers the wavelength range of $3000-24800$~\AA, and was taken simultaneously
in three arms (UVB, VIS and NIR) using slit widths of
1.0, 0.9 and 0.9$''$ for UVB, VIS and NIR respectively. The nominal values of
the spectral resolution for these configurations are $R =\lambda / \Delta \lambda (\text{FWHM}) = 5100$, 8800 and 5100, respectively.

X-shooter spectroscopy started on 2011-10-09  05:55:49 UT, 8.52 hr after the
trigger \citep{2011GCN..12431...1W}, and was performed at a median seeing of
$1.1''$ and an average airmass of 1.04. We obtained five exposures with a total
integration time of 8775~s in the UVB and VIS arm, and 14 exposures (total
exposure of 8400~s) in the NIR arm. A second X-shooter epoch was obtained the following night (starting 20.10 hr after the trigger, on 2011-10-10 04:24:25 UT) with a total integration time 7897 s in UVB and VIS, and 7200 s in NIR. All frames were reduced separately using
the ESO X-shooter pipeline \texttt{v2.0.0} \citep{2010SPIE.7737E..56M}, and the
resulting frames were stacked per night with a weighting based on the signal-to-noise ratio (SNR) of the afterglow detection. Flux calibration was performed against the
spectro-photometric standard GD71 observed starting on 2011-10-08 08:58:03 UT.
The one-dimensional spectra were extracted using optimal, variance weighting
\citep{1986PASP...98..609H}. The optimal extraction and the stacking of science 
frames were done by a custom-made script developed for this purpose. Unless stated explicitly (as in Section \ref{sec_finestructure}), we base our results on the combined spectrum of the first night of X-shooter observations, because of the superior SNR.

The complete VIS and NIR part of the X-shooter spectrum, and the measured absorption lines, are shown in Appendix~\ref{xsh-appendix}.

\subsection{GMOS spectroscopy}

We observed the afterglow of GRB\,111008A using the Gemini Multi-Object
Spectrograph (GMOS) on the Gemini-South telescope\footnote{For an overview of Gemini afterglow spectra see: \url{http://grbspecdb.ucolick.org/}}.
After identification of the optical afterglow \citep{2011GCN..12429...1L}, a series of four exposures with a total integration time of 2400 s were taken, using the R400 grism and a 1.0$''$ slit width, starting at 2011-11-09 03:29:29.7 UT.

The four exposures were taken using dithers in both the dispersion direction (50 \AA) and the spatial direction (along the slit)
to sample over the chip gaps and regions affected by amplifier location.
We reduced the data using the Gemini GMOS reduction package (version 1.11) within IRAF,
combining the four exposures after extraction, resulting in a wavelength range of $\sim3900 - 8170$ \AA. The resolution of the exposures is $R\sim870$.

\subsection{GROND photometry}\label{GrondPhotometry}

The Gamma-Ray burst Optical/Near-infrared Detector (GROND;
\citealp{2008PASP..120..405G}) at the 2.2m MPI/ESO telescope on the La
Silla observatory started photometric observations of the field of
GRB~111008A on 2011-11-09, 04:38:45 UT, which is 6.43~hr after the BAT
trigger. Simultaneous photometry in seven optical/NIR filters (similar
to the SDSS $g'r'i'z'$ and the 2MASS $JHK_{\rm{s}}$ bands) was
obtained continuously until 09:02:10 UT. GROND data were reduced in a
standard fashion \citep{2008ApJ...685..376K}, using a custom pipeline
written in IRAF/pyraf. The photometric solution was obtained by using
magnitudes of stars in an SDSS field \citep{2011ApJS..193...29A}
observed directly after the GRB field under photometric conditions in
the case of the $g'r'i'z'$ filters, and via the magnitudes of 2MASS
stars in $JHK_{\rm{s}}$ \citep{2006AJ....131.1163S}. Based on the
scatter of individual calibration stars, we estimate the absolute
accuracy of our photometry to be 4\% in $g'r'i'z'$, 5\% in $J/H$ and
8\% in $K_{\rm{s}}$.

\begin{figure*}
\centering
\includegraphics[width=0.98\textwidth]{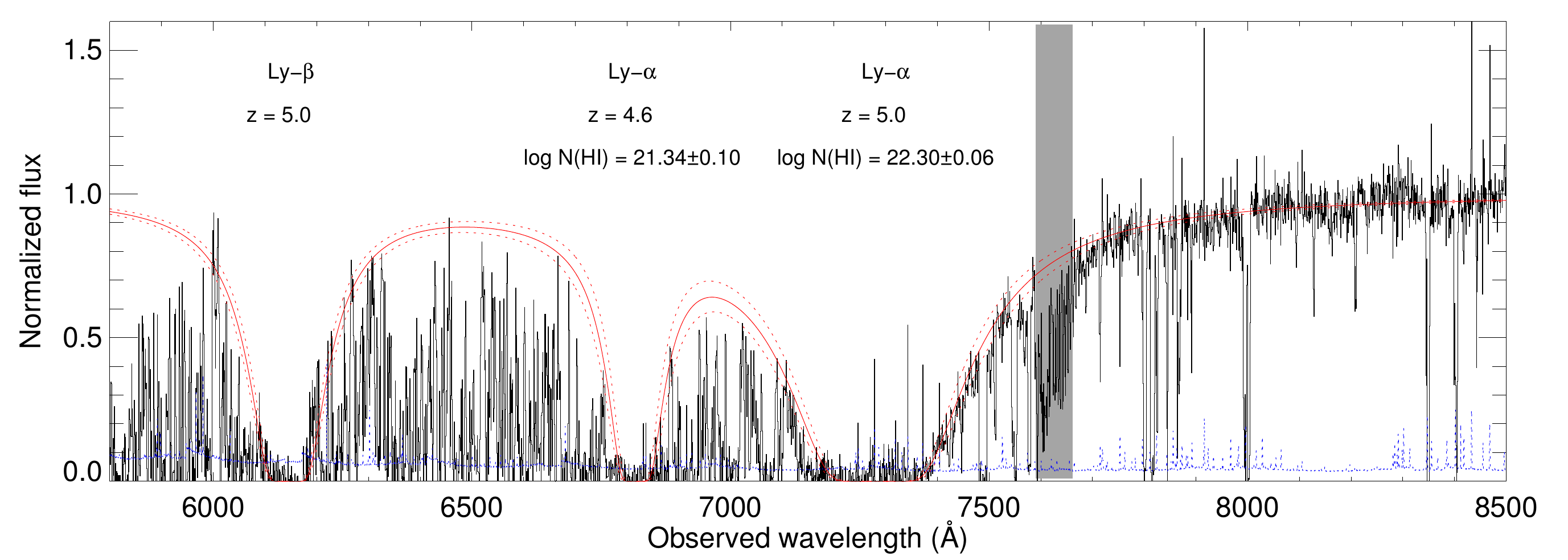}
\caption{Excerpt of the VIS spectrum showing absorption from Ly-$\alpha$ and Ly-$\beta$ of the host galaxy of GRB\,111008A ($z=5.0$) and Ly-$\alpha$ of the strong intervening system at $z=4.6$. The red line displays a Voigt-profile fit to the DLAs Ly-$\alpha$ and Ly-$\beta$ lines. The dashed line shows the 1$\sigma$ error on the fit result, and the dotted line shows the error spectrum. The shaded gray indicates a region with strong telluric contamination. \label{fig:DLA}}
\end{figure*}

\section{The GRB host absorber}\label{sec_metallicity}

In Fig.~\ref{fig:DLA} we provide an excerpt of the VIS spectrum (from X-shooter), which has been rebinned by a factor of 4 for graphical reasons. The VIS spectrum exhibits a very strong DLA, which is identified as absorption within the host galaxy, providing the redshift of the GRB. The corresponding Ly$\beta$ line is also detected. Shown in red 
is a Voigt-profile fit to the DLA and Ly$\beta$ lines corresponding to an
\ion{H}{1} column density of $\log{N/\mathrm{cm^{-2}}} = 22.30\pm0.06$.
Also seen in the VIS spectrum is an intervening DLA at redshift $z=4.6$
which will be discussed in Section~\ref{sec_intervening}. When determining the \ion{H}{1} column density of the two absorbers, we fix the redshift to the value measured from the metal line fit (see Section \ref{section:chempos}), and next we determine $\log N($\ion{H}{1}$)$ subjectively by looking at the red wing of the Lyman-$\alpha$ and Lyman-$\beta$ profiles. The error reflects what we consider the acceptable range of possible column densities.

\subsection{The chemical composition}
\label{section:chempos}

To determine abundances of the atomic ground state levels we proceed with Voigt-profile fitting of the X-shooter spectrum\footnote{The absorption lines are fitted with VPFIT version 10.0: \url{http://www.ast.cam.ac.uk/~rfc/vpfit.html}}. We find that a
model with two absorption components per transition is sufficient to fit
the low-ionization absorption lines.  The spectral resolution is set to the nominal
values from the X-shooter manual (a velocity resolution of 34 km s$^{-1}$ for
VIS and 57 km s$^{-1}$ for NIR, both full-width-at-half-maximum). The choice of spectral resolution and the placement of the continuum are unlikely to change the logarithm of the column densities by more than 0.15 dex.

\begin{figure}
\centering
\includegraphics[width=0.98\linewidth]{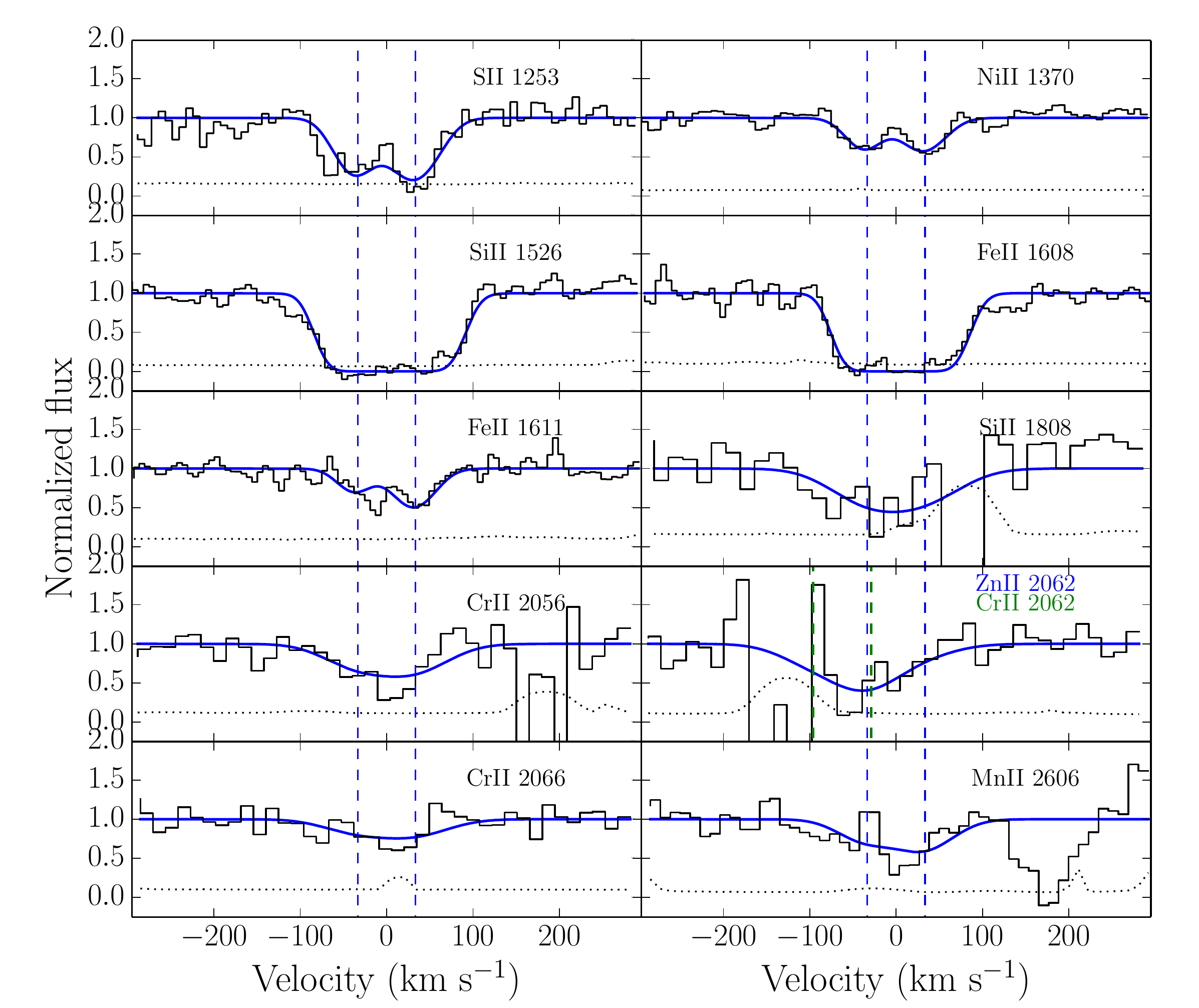}
\caption{The fitted absorption lines from the host galaxy at $z=5.0$. The black line is the spectrum, the blue line is the fitting model and the dotted line is the error spectrum. The centers of the absorption line components are marked with dashed vertical lines.}
\label{Lines}
\end{figure}

The two Voigt-profiles have Doppler parameters of $b=20.9\pm2.3$ km s$^{-1}$ and $b=27.7\pm 2.3$ km s$^{-1}$, and
redshifts of $4.99005 \pm 0.00007$ and $4.99142 \pm 0.00006$, respectively (these are the precise measured redshifts for the 
two absorption components in the GRB host absorption system. For brevity we will refer to the GRB host redshift as 5.0 in the remaining parts of the article). We linked the Doppler parameters for the different atoms assuming turbulent broadening, which is a standard procedure for low-ionization absorption lines \citep{2005ARA&A..43..861W}. We also linked the redshifts for each of the absorption components. Fig.~\ref{Lines} shows the fit of the line transitions for \ion{Si}{2}, \ion{S}{2}, \ion{Cr}{2}, \ion{Mn}{2}, \ion{Fe}{2}, \ion{Ni}{2} and
\ion{Zn}{2}.

The derived metal column densities for the sum of the two components are
provided in Table~\ref{table:metallicity}. Note that this table also includes the column densities in excited levels (from Section~\ref{sec_finestructure}). The best determined column
densities are Ni and Fe, since they are constrained by relatively weak
transitions (\ion{Ni}{2} $\lambda$1370 and \ion{Fe}{2} $\lambda$1611) located in regions with a good SNR. For \ion{Fe}{2} $\lambda$1611 the measured spectrum exhibits excess absorption (at $\simeq-15$ km s$^{-1}$) compared to the fit model. It is unclear whether this feature is due to an additional absorption component or a systematic error. If it is due to an additional absorption component our column density of \ion{Fe}{2} is slightly underestimated by $\lesssim $15\%.

For the \ion{S}{2} $\lambda$1253 transition the SNR is also high, and the measurement is reliable even though this line is mildly saturated (see a detailed explanation in Appendix~\ref{S_saturation}). For \ion{Cr}{2} we were able to estimate the column density with \ion{Cr}{2} $\lambda$2056 and \ion{Cr}{2} $\lambda$2066.

\subsubsection{Systematic errors in the fitting of Si, Mn and Zn for the GRB host}\label{sys-errors}

The estimate of the column density of Si is unreliable because it is estimated using two lines, where one of them is saturated (\ion{Si}{2} $\lambda$1526) and the other is affected by a sky-line (\ion{Si}{2} $\lambda$1808). We therefore only report a lower limit on the column density based on \ion{Si}{2} $\lambda$1808. The estimated column density of Zn is also uncertain, since the blue component of the \ion{Zn}{2} $\lambda$2062 transition is blended with \ion{Cr}{2}
$\lambda$2062\footnote{In our estimate we assumed that $\log N/$cm$^{-2}$ for the blue component of Zn is 0.1 dex lower than for the red component, since this is what we see for \ion{Ni}{2} $\lambda$1370.}. Visually, the model fit of \ion{Mn}{2} $\lambda$2606 does not convincingly fit the data (and $\chi^2/\text{d.o.f.}=4.55$ for this transition, which also indicates a bad fit), so our measurement of the Mn column density is unreliable.

In the last column of Table~\ref{table:metallicity} these caveats are summarized. Since the derived column densities of Si, Mn and Zn are unreliable, we will not draw any conclusions based on them.

\begin{table*}
\caption{The abundance pattern of the transitions (including excited levels) in the GRB 111008A host galaxy and for the intervening absorber. 
Lower limits are at a $2\sigma$ level. Abundances from the solar photosphere from \citet{2009ARA&A..47..481A} were used as reference.}
\label{table:metallicity}
\centering 
\begin{tabular}{lccll}
\hline\hline 
Element & $\log N(X) / $cm$^{-2}$ & $[$X$/$H$]$& Lines used in fit&Note\\ \hline 
&  GRB host\\
H&$22.30\pm0.06$&-& Ly $\alpha$, Ly $\beta$\\
Si &$>15.86 $ (2$\sigma$)& $ >-1.96$ (2$\sigma$)& \ion{Si}{2} $\lambda$1808, \ion{Si}{2} $\lambda$1526&Lines are saturated\\
S & $15.71 \pm 0.09 $ & $ -1.70 \pm 0.10$&\ion{S}{2} $\lambda$1253\\
Cr & $14.17 \pm 0.09 $ &$ -1.76 \pm 0.11$&\ion{Cr}{2}
$\lambda$2056, \ion{Cr}{2}
$\lambda$2062, \ion{Cr}{2}
$\lambda$2066\\
Mn & $13.72 \pm 0.08 $ & $ -2.01 \pm 0.10$&\ion{Mn}{2} $\lambda$2606 & Fit is not convincing\\
Fe & $16.05 \pm 0.05 $ & $-1.74 \pm 0.08$   &\ion{Fe}{2}
$\lambda$1608, \ion{Fe}{2}
$\lambda$1611\\
Ni&$14.89 \pm 0.18 $ & $-1.64 \pm 0.19$&\ion{Ni}{2} $\lambda$1370\\
Zn & $13.28 \pm 0.21 $ & $ -1.58 \pm 0.21$&\ion{Zn}{2} $\lambda$2062&Blended with \ion{Cr}{2} $\lambda$2062\\
\hline 
& $z=4.6$ system\\ 
H&$21.34\pm 0.10$&-&Ly $\alpha$\\
Si& $>14.91$ (2$\sigma$)& $>-2.15$ (2$\sigma$)& \ion{Si}{2} $\lambda$1526&Lines are saturated\\
Fe& $15.23\pm 0.15$& $-1.61\pm 0.17$&\ion{Fe}{2} $\lambda$1608, \ion{Fe}{2} $\lambda$1611\\
Ni& $14.23 \pm 0.08$ & $-1.33\pm 0.12$&\ion{Ni}{2} $\lambda$1370, \ion{Ni}{2} $\lambda$1741\\
\hline\hline
\end{tabular}
\end{table*}

\subsection{Fine-structure lines}\label{sec_finestructure}

At $z=5.0$ we detect lines that arise from the following fine-structure and metastable levels: C\,{\sc ii} $^2P^\circ_{3/2}$ ($^{**}$), O\,{\sc i} $^3P^\circ_{0}$ ($^{**}$), Si\,{\sc ii} $^2P^\circ_{3/2}$ ($^{*}$), Fe\,{\sc ii} ($^6D_{7/2}$, $^6D_{5/2}$, $^6D_{1/2}$, $^4F_{9/2}$, $^4F_{5/2}$, $^4D_{7/2}$, $^4D_{5/2}$, $^4D_{3/2}$) and Ni\,{\sc ii} $^4F_{9/2}$. To date, this is the highest redshift at which lines of these excited states of \ion{Fe}{2} and \ion{Ni}{2} have been detected. Lines from these excited levels are expected in GRB host galaxies and often detected at lower redshifts, but due to their relative weakness the SNR required to detect them is often not reached at high redshift. Lines from Fe\,{\sc ii} $^6D_{3/2}$ and $^4F_{7/2}$ are not clearly detected due to their unfortunate placing either outside atmospheric windows or because they are severely affected by telluric lines. The Gemini/GMOS and X-shooter data together cover a time span from 5 to 40 hours after the burst (observers frame), making the data set suited to look for variability in lines from excited levels. Line variability is expected, because the fine-structure and metastable levels are populated through indirect UV-pumping by the GRB afterglow \citep{prochaska2006,vreeswijk2007}. The lines from excited states and their corresponding ground states that fall in the spectral region covered both by GMOS and X-shooter are all likely saturated, therefore we compare their rest-frame equivalent width ($\mathrm{EW}_\mathrm{rest}$), see Table~\ref{table:finestruc}. The values of $\mathrm{EW}_\mathrm{rest}$ for the lines and line blends (including both resonance and excited states) at the top of Table~\ref{table:finestruc} are constant in time within $2\sigma$, except Si\,{\sc ii}$^{*}\,\lambda1309$. However, the temporal variation of this line does not match that of the stronger Si\,{\sc ii}$^{*}\,\lambda1264$ lines, which it should follow. One possible explanation for this is that the uncertainty on Si\,{\sc ii}$^{*}\,\lambda1309$ are probably underestimated, especially in the last epoch.

The lines from the excited states of Fe\,{\sc ii} and Ni\,{\sc ii}, which are not covered by GMOS, are weaker (and not saturated) and can be fitted with Voigt profiles in the X-shooter spectra. We couple $z$ and $b$ and fit one component to the lines of these levels in the first X-shooter epoch. The redshift is consistent with the red component of the resonance lines. With the obtained $z$ and $b$ kept fixed, this fit is repeated on the second X-shooter epoch, though none of the lines are clearly detected. Therefore, we report 2$\sigma$ upper limits on the population of these excited states (see Table~\ref{table:finestruc}). Here again we see no evidence for time variation for an individual excited state. We note, that the upper limits on the derived column densities are larger in the second X-shooter epoch compared to the first X-shooter epoch, since the signal-to-noise is lower in the second epoch.

From the measurements on the first X-shooter spectrum we conclude that the column density of the metastable level Fe\,{\sc ii}\,$^4F_{9/2}$ is, despite its higher energy, as high as that of the first excited state Fe\,{\sc ii}\,$^6D_{7/2}$ and that of all other, lower energy Fe\,{\sc ii} fine-structure states. Furthermore, Ni\,{\sc ii}\,$^4F_{9/2}$ is more populated than the Ni\,{\sc ii} ground state. This situation is typically the result of population by indirect radiative pumping. The Ni\,{\sc ii}\,$^4F_{9/2}$ population is expected to peak much later than the Fe\,{\sc ii} fine-structure states (after $\sim 2-10$ hr post burst in the rest frame, depending on light curve shape, see also \citealt{hartoog2013}), which is consistent with our time of observations. Although a detailed model of the excitation has not been carried out in this paper, and despite the fact that we do not have evidence for variability in individual lines, our observations are consistent with the UV-pumping scenario, which confirms the $z=5.0$ DLA as the host galaxy.

\begin{table*}
\caption{Measurements of lines from fine-structure and metastable states at the host-galaxy redshift ($z=5.0$) in three different epochs. The limits are at a $2\sigma$ confidence level.}
\label{table:finestruc}
\centering 
\begin{tabular}{llccc}
\hline
\multicolumn{2}{l}{Time of mid exposure$^a$  (hr)}&6.27    	& 10.15 					& 34.82\\
	&	& GMOS		&	X-shooter	 2011-11-09	& X-shooter 2011-11-10\\
\hline
Ion&Line (blend)&\multicolumn{3}{c}{$\mathrm{EW}_{\mathrm{rest}}$ (\AA) of saturated lines and blends}\\
\hline
C\,{\sc ii}\,+\,C\,{\sc ii}$^{*}$  	&	$\lambda	1334+\lambda1335	$	&$2.05\pm0.33$&$	1.94	\pm	0.04	$&$	1.77	\pm	0.12	$\\
O\,{\sc i} 					&	$\lambda	1302	$	&$1.02\pm0.02$&$	1.05	\pm	0.03	$&$	0.93	\pm	0.05	$\\
O\,{\sc i}$^{**}$ 			&	$\lambda	1306	$	&$0.62\pm0.17$&$	0.36	\pm	0.03	$&$	0.42	\pm	0.07	$\\
S\,{\sc ii}\,+\,Si\,{\sc ii} 		&	$\lambda	1259+\lambda1260	$	&$1.58\pm0.03$&$	1.64	\pm	0.03	$&$	1.79	\pm	0.14	$\\
Si\,{\sc ii}$^{*}$ 			&	$\lambda	1264+\lambda1265$		&$1.15\pm0.03$&$	1.05	\pm	0.03	$&$	1.12	\pm	0.10	$\\
Si\,{\sc ii}$^{*}$ 			&	$\lambda	1309	$	&$0.37\pm0.03$&$	0.40	\pm	0.02	$&$	0.14	\pm	0.05	$\\
\hline
Ion&Level&\multicolumn{3}{c}{Column densities $\log N(X)/\mathrm{cm}^{-2}$ from line fits}\\
\hline
 Fe\,{\sc ii}  	&$^6D_{9/2}$ (ground)			&	$^b$		&	$15.72\pm0.08^{c}$		&$<16.14$ \\
			&$^6D_{7/2}$ 					&	$^b$		&	$14.63\pm0.06$		&$<14.74$\\
  			&$^6D_{5/2}$					&	$^b$		&	$14.53\pm0.07$		&$<14.80$\\
 			&$^6D_{1/2}$					&	$^b$		&	$13.71\pm0.11$		&--\\
			 &$^4F_{9/2}$					&	$^b$		&	$14.66\pm0.12$		&--\\
			 &$^4F_{5/2}$					&	$^b$		&	$14.09\pm0.24$		&--\\
			 &$^4D_{7/2}$					&	$^b$		&	$13.66\pm 0.05$		&$<13.68$\\
			 &$^4D_{5/2}$					&	$^b$		&	$13.34\pm0.08$		&$<13.91$\\
			 &$^4D_{3/2}$					&	$^b$		&	$12.95\pm0.12$		&$<14.01$ \\
Ni\,{\sc ii}           &$^2D_{5/2}$	(ground)			&$14.62 \pm 0.77 $		&	$14.34\pm0.05$		& $14.33 \pm 0.20 $
\\
			 &$^4F_{9/2}$					&	$^b$	&	$14.73\pm0.26$		& -\\
\hline
\multicolumn{5}{l}{$^a$Time since burst.}\\
\multicolumn{5}{l}{$\phantom{^a}$Not taking into account that we use an average spectrum weighted by the SNR of a fading source.}\\
\multicolumn{5}{l}{$^b$All lines are outside the spectral range of GMOS.}\\
\multicolumn{5}{l}{$^c$This is only the red component, which coincides with the position of the fine-structure lines.}\\
\hline
\end{tabular}
\end{table*}

\begin{figure}
\centering
\includegraphics[width=0.98\linewidth]{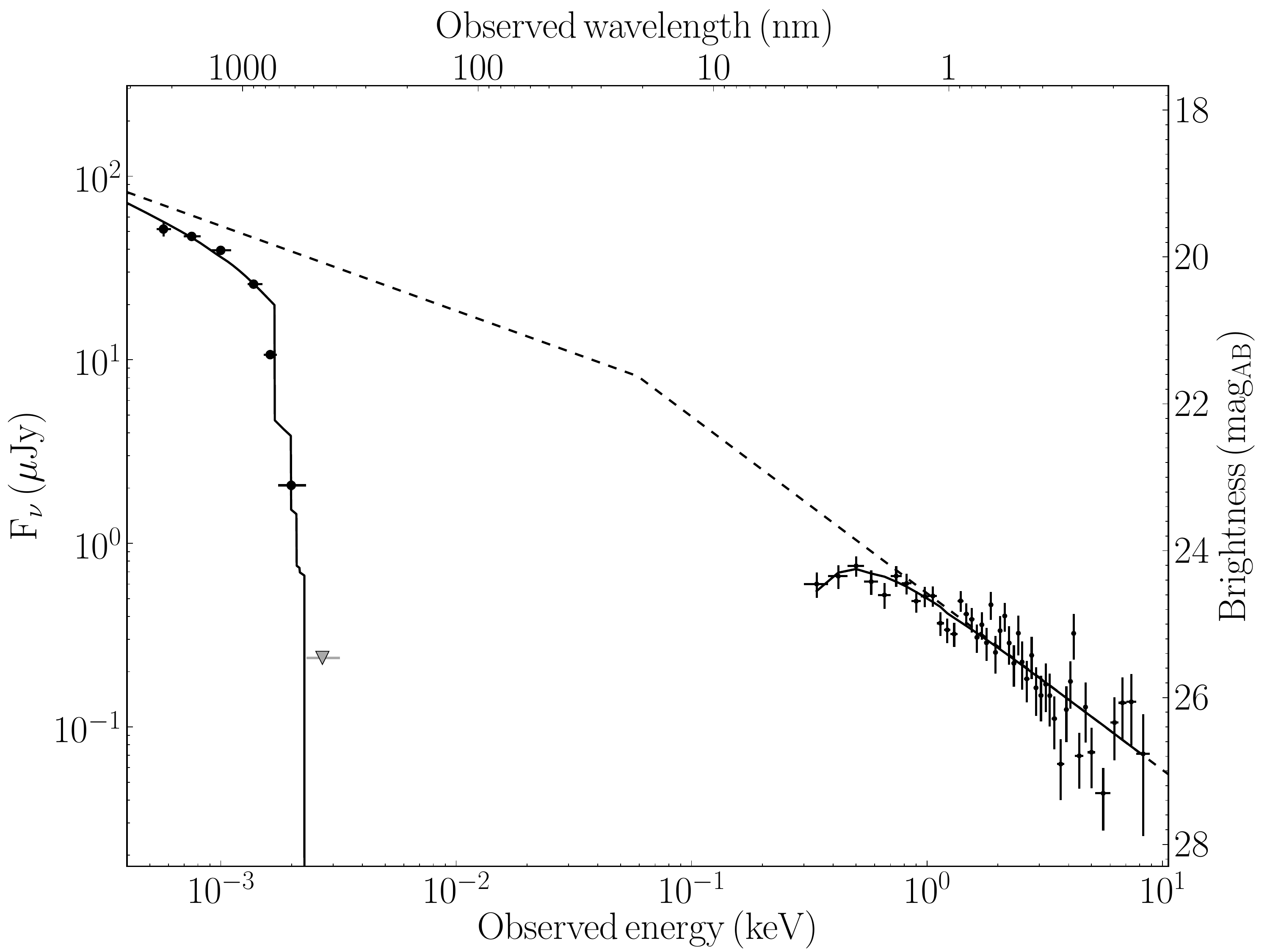}
\caption{NIR-to-X-ray spectral energy distribution and model for the afterglow of GRB~111008A at an observed time of 35~ks after the GRB trigger. GROND photometry and the $g'$ band upper limit are shown in larger black circles and a grey downward triangle, respectively. \textit{Swift}/XRT X-ray data are plotted in smaller black dots. The best-fit model including gas and dust absorption is shown with solid lines, while the dashed line illustrates the underlying synchrotron emission. X-ray data have been binned to yield a S$/$N of at least 8 to enhance clarity. The $g'r'i'$-band photometry is not fitted, because these filters are located or extend bluewards of the Ly$\alpha$ transition.}
\label{GROND_SWIFT}
\end{figure}

\section{Determining the dust extinction}\label{sec_grond}

We retrieved the X-ray spectrum of the afterglow of GRB~111008A from
the \textit{Swift}/XRT repository \citep{2007A&A...469..379E,
2009MNRAS.397.1177E} and fit it together with the GROND broad-band
photometry in a standard manner \citep[see e.g.,][for
details]{2013A&A...557A..18K}. The fit is shown in Fig.~\ref{GROND_SWIFT}. We use synchrotron emission
models, reddened by extinction laws from the Local Group
\citep{1992ApJ...395..130P}, and fit them to the available
data. The absorption of soft X-rays is modeled with two absorbers at
solar metallicity, one in the Galaxy (\citealt{2005A&A...440..775K}, see also \citealt{2013MNRAS.431..394W}), the
other at the GRB redshift. The fit is performed with data at a mean
photon arrival time of 35~ks after the trigger. AB magnitudes of the
afterglow at this epoch in the different filters are $g' >
25.5\,\rm{mag}$, $r' = 23.11 \pm 0.07\,\rm{mag}$,  $i' = 21.33 \pm
0.05\,\rm{mag}$, $z' = 20.37 \pm 0.05\,\rm{mag}$, $J = 19.91 \pm 0.06\,\rm{mag}$,
$H = 19.72 \pm 0.06\,\rm{mag}$, $K_{\rm{s}} = 19.62 \pm 0.10\,\rm{mag}$.
We note, that the $g'r'i'$-band data are not part of the fit, because they
are located bluewards of the Ly$\alpha$ transition. For the $z'$-filter, the strong
absorption lines from the GRB-DLA and strong intervening system reduce the
observed flux significantly. We use the X-shooter spectrum to estimate their effect
on the $z'$-band measurement, and find that the continuum emission is approximately
$7\pm1\,$\% above what is measured with GROND\@. This correction factor is applied in the 
following. The effect on the $JHK_{\rm{s}}$ magnitudes is $\lesssim 4\,$\%.

The data are well fitted ($\chi^2 = 1.6$ in the optical/NIR, and Cash-statistic = 385
in the X-ray energy range, for a total of 362 degrees of freedom) with a
broken power-law with a low-energy spectral index $\beta_1 = 0.46 \pm 0.06$
and a small amount of reddening in an SMC-like (Small Magellanic Cloud) extinction law
($E_{B-V} = 0.037 \pm 0.012\,\rm{mag}$, corresponding to a visual extinction of
$A_V = 0.11 \pm 0.04\,\rm{mag}$). The high-energy spectral index $\beta_2$ is
tied to $\beta_1$ through $\beta_2=\beta_1+0.5$ as expected for synchrotron
emission (predicted by \citealt{1998ApJ...497L..17S}, and observed by \citealt{2011A&A...526A..30G,2011A&A...532A.143Z} and \citealt{2013MNRAS.432.1231C}), and the best fit soft X-ray absorption at $z = 5.0$ corresponds to
$N(\rm{H})_{\rm{X}} = (1.8\pm0.4)\times10^{22}\,\rm{cm^{-2}}$. Because of the
small amount of reddening, an LMC (Large Magellanic Cloud) dust model provides a reasonable description
of the optical/NIR data as well ($\chi^2 = 3.8$), and yields
similar values for all parameters within the errors. A fit with a MW-dust model is significantly
worse ($\chi^2 = 9.4$) because of the lack of a 2175~\AA~dust feature in our data.

\begin{figure}
\centering
\includegraphics[width=0.98\linewidth]{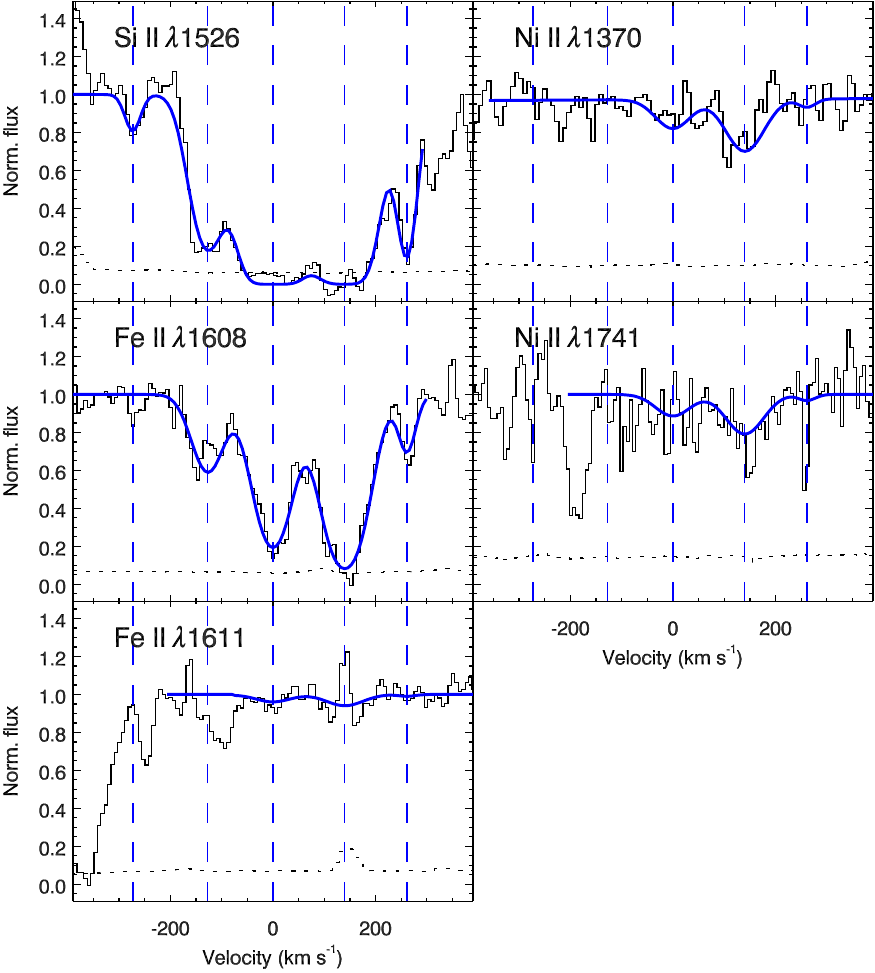}
\caption{The most constraining metal lines from the intervening absorber at $z = 4.6$. The solid profiles are resulting Voigt profile fits with five absorption components (the dashed vertical lines show the velocity of the absorption components). The dotted line shows the error spectrum.}
\label{FineStructure}
\end{figure}

\section{The intervening system at $\lowercase{z=4.6}$} \label{sec_intervening}

In the line of sight towards the afterglow of GRB\,111008A an additional DLA system is detected at $z = 4.6$, with a neutral hydrogen column density of $\log N(\mathrm{HI})/\text{cm}^{-2}=21.34\pm0.10$ (Fig.~\ref{fig:DLA}). A number of strong metal absorption lines are detected from this system, including C{\,\sc iv}, Mg{\,\sc i}, Mg{\,\sc ii}, Si{\,\sc ii}, Si{\,\sc iv}, Fe{\,\sc ii} and Ni{\,\sc ii}. We note that the $\mathrm{EW}_\mathrm{rest}$ of Si{\,\sc ii}\,$\lambda1526$ is the second highest ever detected for a DLA at $z>4$ ($\mathrm{EW}_\mathrm{rest}=2.30\pm0.02$ \AA{}), indicating a relatively high metallicity for this redshift \citep{2008ApJ...672...59P, 2012ApJ...755...89R}.

The low as well as the high ionization lines show multiple components. Due to their saturation, we do not include the Si{\,\sc iv} and C{\,\sc iv} lines in our analysis. The absorption lines from singly ionized species are fitted with Voigt profiles, in which we assume that the different ions have the same velocity structure; i.e. equal $z$ and $b$-parameters per velocity component. In Fig.~\ref{FineStructure} we show the best fit for the most constraining absorption lines, which needs five components for the strongest lines. In the Voigt profile fitting we have included more lines than shown in the figure, most of which are visually absent, but we have avoided regions with strong telluric contamination. Including both strong and weak lines, predominantly of Fe{\,\sc ii}, helps to constrain the $b$ parameters. For the different absorption components the Voigt profile fitting results in the following $b$-parameters ($z$-values): $5.0\pm 27.7$ km s$^{-1}$ ($4.60337 \pm 0.00009$), $36.4\pm 2.8$ km s$^{-1}$ ($4.60611 \pm 0.00004$), $41.8\pm 2.9$ km s$^{-1}$ ($4.60848 \pm 0.00003$), $45.2\pm 2.3$ km s$^{-1}$ ($4.61110 \pm 0.00003$) and $10.7\pm 2.0$ km s$^{-1}$ ($4.61337 \pm 0.00002$). Outside this section we will refer to the redshift of the intervening absorber as $z=4.6$.

The total column densities of Si{\,\sc ii}, Fe{\,\sc ii} and Ni{\,\sc ii} can be constrained, and are summarized in Table~\ref{table:metallicity}. The column densities of Fe{\,\sc ii} and Ni{\,\sc ii} are robust because they are determined from optically thin lines. The fit for Si{\,\sc ii} is mainly determined by the saturated \ion{Si}{2}\,$\lambda$1526 transition, as \ion{Si}{2}\,$\lambda$1808 is in a low SNR region, so we conservatively report a $2\sigma$ lower limit of $\log N\left(\mathrm{Si\,\sc II}\right)/$cm$^{-2} > 14.91$ based on the equivalent width of this line.

The conservative lower limit on the metallicity based on the equivalent width of Si{\,\sc ii}\,$\lambda1526$ is 1\% solar. The value of $[$Ni$/$H$]$ indicates a metallicity of 3--6\% solar, and $[$Fe$/$H$]$ corresponds to 2--4\% solar. For QSO and GRB DLAs at these metallicities \citep[e.g.,][]{Dessauges06, 2007ApJ...666..267P, 2003ApJ...585..638S} these elements (especially Ni and Fe) are often depleted onto dust grains, which implies that the true metallicity might be (much) higher. Generally, the ratio $[$Zn$/$Fe$]$, if available, is used to correct for dust depletion, since Zn is expected to be mainly in the gas phase even if a lot of dust is present. Unfortunately, for this intervening absorber we could not reliably constrain the column density of Zn, since \ion{Zn}{2} $\lambda$2026 is coincident with a skyline, and \ion{Zn}{2} $\lambda$2062 is blended with \ion{Cr}{2} $\lambda$2062.

\section{Searching for emission from the host and from the intervening absorber}\label{EmissionSearch}

In the night of August 1st to August 2nd 2013, we obtained deep imaging of the field of GRB~111008A using the ESO/Very Large Telescope and FORS2 \citep{1998Msngr..94....1A} to search for the counterparts of the host and for the intervening DLA in emission (as it has been done for e.g. GRB 070721B, see \citealt{2012A&A...546A..20S}). Our FORS2 observations consisted of 31 dithered exposures of 3 minutes integration time each in the $z_{\rm{Gunn}}$-band filter which is centered around $9100\,$\AA. The data were processed and calibrated in a similar way to the GROND imaging data (see Section \ref{GrondPhotometry}). The stacked FORS2 image has a full-width at half maximum of the stellar PSF of 0.9$''$ and reaches a 2$\sigma$ depth of a $z$-band magnitude of $26\,\rm{mag_{AB}}$.

We do not detect emission centered at the position of the optical transient, and we set an upper limit of $z > 25.6\,\rm{mag_{AB}}$ for the brightness of the GRB host galaxy by measuring the flux at the GRB position in an aperture of size of one FWHM. Especially at the highest redshifts, GRB hosts are faint \citep{2012ApJ...756..187H, 2012ApJ...754...46T} and the lack of a clear counterpart to the GRB-DLA in our imaging is thus not particularly surprising. 

We detect significant emission centered at a projected distance of $0.65\pm0.15''$ to the optical counterpart with an AB magnitude of $z = 25.4\pm0.3$. We consider this a more likely candidate for the counterpart of the intervening DLA at $z=4.6$ than for the $z=5.0$ host system. Also, finding an unrelated galaxy at these flux levels is not unlikely. The probability (estimated following \citealt{2002AJ....123.1111B}) of finding a random field galaxy with $z<25.4\,\rm{mag}$ at this distance to the GRB is approximately 3 percent. An association between the detected source and the DLA could be supported or rejected from further imaging, for example. At $z=4.6$, a galaxy is expected to show a strong Lyman-$\alpha$ break between the $r$ and the $i$-band and no flux transmitted below $5100\,$\AA. 

If associated with the DLA at $z=4.6$ the FORS2 $z$-band measurement yields an absolute magnitude of $M_{\rm{UV}}=-20.9\pm0.3\,\rm{mag}$ at a rest-frame wavelength of $\sim1600\,$\AA, typical of the brightest Lyman break galaxies at this redshift. Compared to the galaxy luminosity function, this magnitude corresponds to $\approx L_{\star}$ \citep{2007ApJ...670..928B}. The measured spatial offset between GRB line of sight and galaxy center would be $4\pm1\,\rm{kpc}$ at $z=4.6$.

Since the absorber has a high metallicity and a high velocity width the chances are that this absorber is part of a massive galaxy \citep{2008ApJ...681..881W}. This would agree with the detection this bright Lyman break galaxy.

\section{Dust-to-metals ratio}\label{DustAndMetals}

The study of the dust-to-metals ratio (\dtm{}) as a function of metallicity holds the potential to probe the dust formation mechanism. If dust is primarily produced by SNe, the \dtm{} is expected to be independent of metallicity \citep{2003MNRAS.343..427M}. If dust grains grow in the ISM, a decline of the \dtm{} is expected at low metallicities \citep{2009ASPC..414..453D, Mattsson12,2012ApJ...752..112H}. The host galaxy of GRB\,111008A is characterized by a variety of metal absorption lines and the afterglow SED is well-calibrated. Thus, in this line of sight we have the opportunity to examine the question of the \dtm{} of the GRB host galaxy through dust extinction along the line of sight.

For a system with a given extinction and metallicity we can calculate the dust-to-metals ratio relative to the Local Group value as
\begin{align}
\textit{DTM} = \frac{1}{\textit{DTM}_\text{LG}}\times \frac{A_V}{N_\text{HI} \times 10^{[\text{M}/\text{HI}]}},
\end{align}
where $\textit{DTM}_\text{LG} \equiv 10^{-21.3}$ mag cm$^2$ is the dust-to-metals ratio in the Local Group \citep{Watson11}. Recently, \citet{2013arXiv1303.1141Z}, using different classes of objects, found that \dtm{} is independent of galaxy type or age, redshift, or metallicity, 
and is very close to the value in the Local Group.

The measured extinction along the sight line of GRB 111008A is affected by the host galaxy as well as the intervening absorber. Assuming that all the extinction is from the host galaxy and that the host galaxy has a metallicity of $[\text{M}/\text{H}]=[\text{S}/\text{H}]$, we find that the host galaxy has
\begin{align}
\textit{DTM}=0.57 \pm 0.26.
\end{align}
We calculated the mean and the error of the $\textit{DTM}$ in a Monte Carlo fashion, where 50000 values for both $A_V$, $\log N_\text{H}$ and $[\text{M}/\text{H}]$ are drawn from normal distributions with $A_V=0.11\pm 0.04$ mag, $\log N_\text{H}/\text{cm}^{-2}=22.30\pm 0.06$ and $[\text{M}/\text{H}]=-1.70 \pm 0.10$. For each triplet of sampled values a $\textit{DTM}$ is calculated, and finally the mean and the standard deviation of the 50000 $\textit{DTM}$-values are computed.

\begin{figure}
\centering
\includegraphics[width=\linewidth]{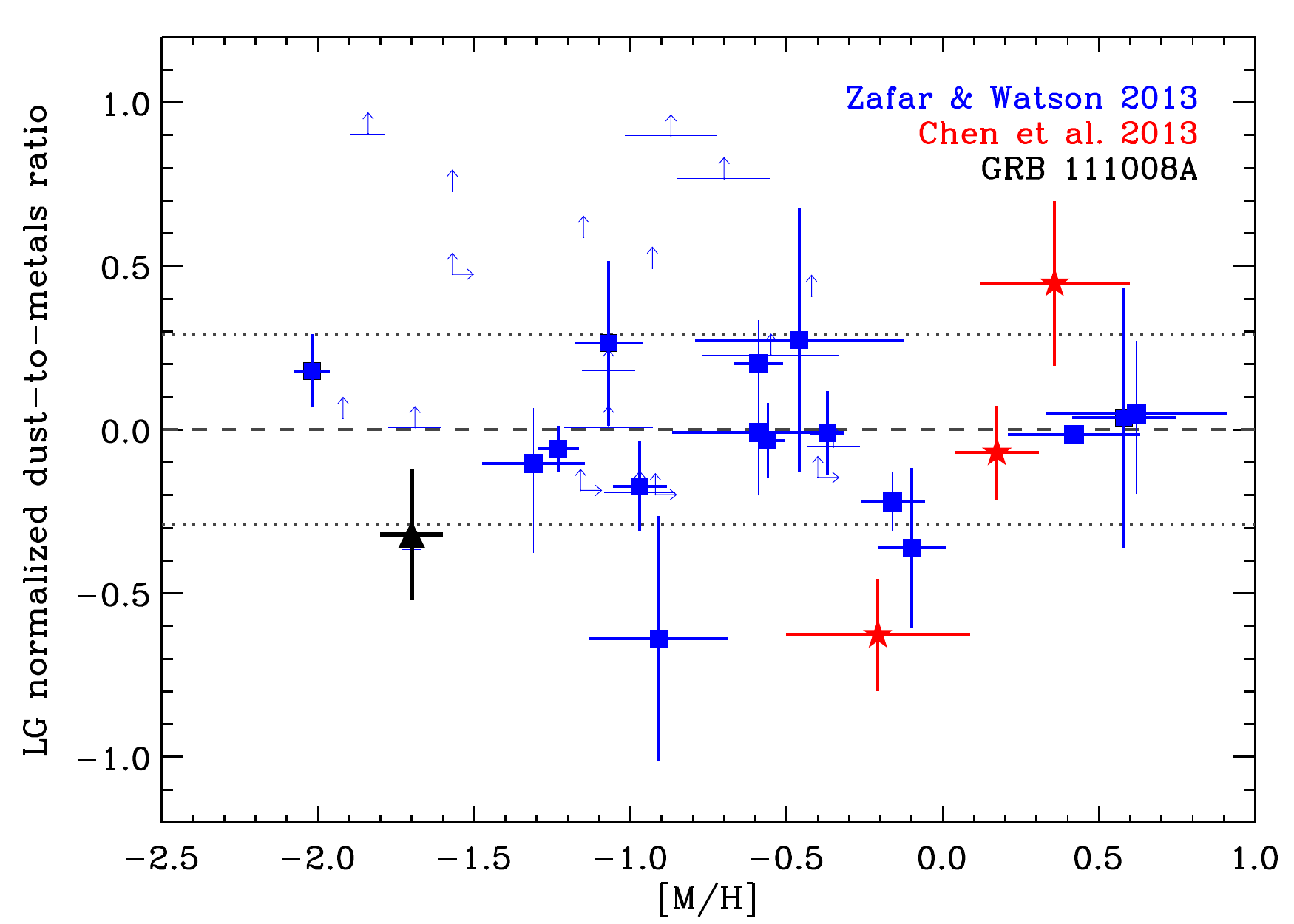}
\caption{
The dust-to-metals ratio versus metallicity.
The dashed line shows the Local Group value, and the dotted lines indicate
the scatter in the Local Group. The host of GRB 111008A, which
is shown as a black triangle, has a dust-to-metals ratio which is
consistent with being equal to or lower than the value in the Local
Group. The other data points are from \citet{2013arXiv1303.1141Z} and \citet{2013arXiv1306.0008C}. }
\label{figure:dtm}
\end{figure}

In Fig. \ref{figure:dtm} we plot the \dtm{} versus the metallicity for the GRB 111008A host with GRB-DLAs, QSO-DLAs, nearby lensed galaxies, and with the value
found in the Local Group (see \citealt{2013arXiv1303.1141Z} and \citealt{2013arXiv1306.0008C} for details). Given the uncertainty in $A_V$, the host has a \dtm{}, which -- within a $1\sigma$ error bar -- is consistent with the \dtm{} observed in the Local Group. Our data is also consistent with the scenario where the host has a much lower \dtm{} than the Local Group, especially if part of the extinction along the line of sight is due to the intervening absorber at $z=4.6$.

\begin{figure}
\centering
\includegraphics[width=\linewidth]{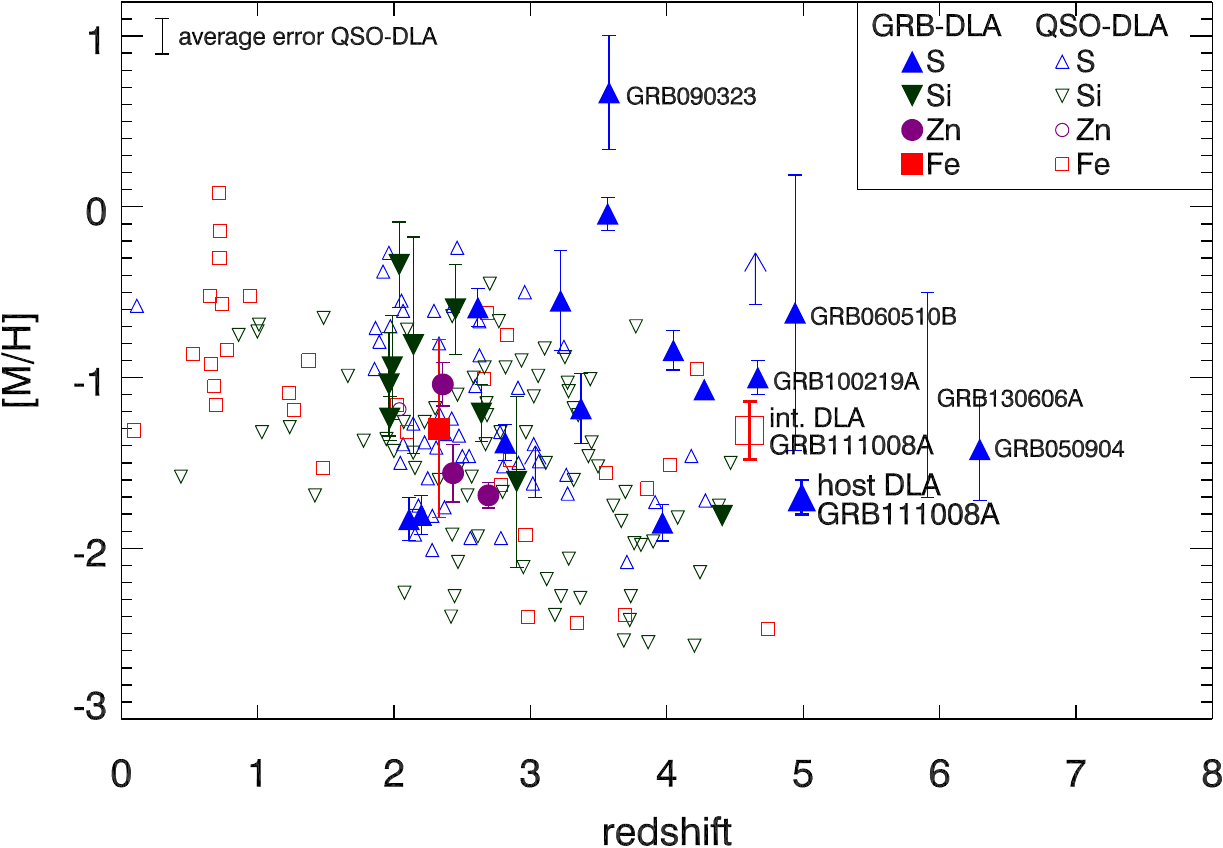}
\caption{Absorption-line based metallicities $[\text{M}/\text{H}]$ as a function of redshift for GRB-DLAs (filled symbols) and QSO-DLAs (open symbols). The figure is adapted from \citet{2012ApJ...755...89R}, with added GRB-DLAs from the following references: \citet{2003ApJ...585..638S,2004A&A...419..927V,2005ApJ...624..853F,2006Natur.440..184K,2006ApJ...652.1011W,2006ApJ...642..979B,2007ApJS..168..231P,2007ApJ...663L..57P,2007ApJ...671..272C,2008A&A...489...37T,2009ApJ...691L..27P,2010A&A...522A..20D,2010A&A...523A..36D,2011A&A...525A.113S,2011MNRAS.412.2229D,2012MSAIS..21..206D,2013MNRAS.428.3590T,2013A&A...557A..18K,2013ApJ...774...26C,2014arXiv1402.4026D}
(For the measurement of the metallicity of GRB~130606A from \citet{2013ApJ...774...26C} the lower limit is based on Si and the upper limit is based on S. We have indicated this by the colors of the error bar belonging to this data point).
Following \citet{2012ApJ...755...89R}, we apply $[\text{M}/\text{H}]=[\text{Fe}/\text{H}]+0.3$ in cases where the metal is iron, and no less-refractory element abundance is measured (this correction is also applied to the intervening absorber in the GRB 111008A sight line).}
\label{Comparison}
\end{figure}

\begin{figure}
\centering
\includegraphics[width=\linewidth]{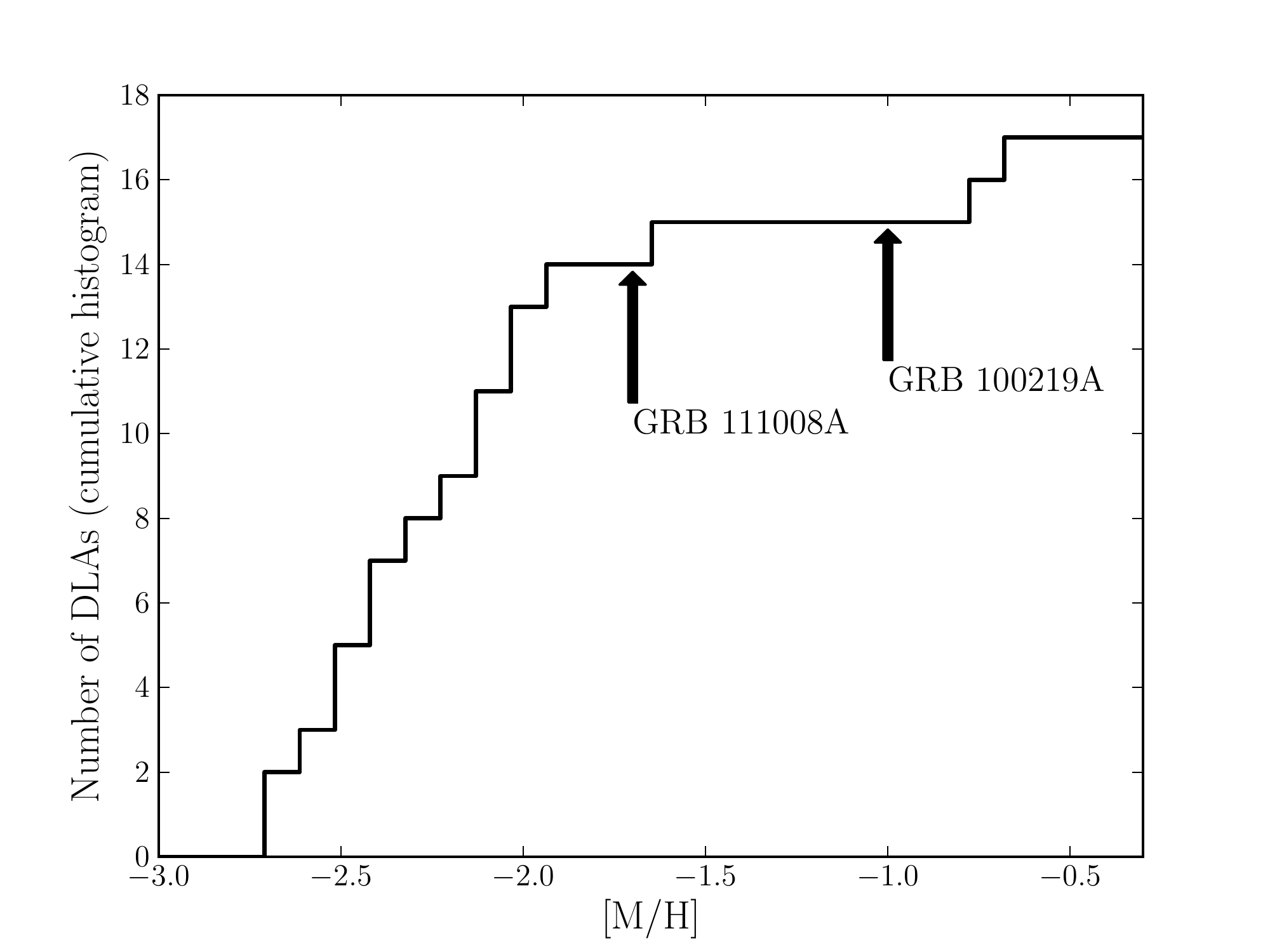}
\caption{This figure compares the metallicity of the two high-redshift GRBs 100219A and 111008A with the 17 QSO-DLAs at $3.64<z<5.08$ from \citet{2013arXiv1310.6042R}. The two GRB-DLAs are clearly in the upper part of the metallicity distribution of QSO-DLAs, a trend which is also present at lower redshift.}
\label{RafelskiComparison}
\end{figure}

\section{Discussion}\label{sec_interpretation}

\subsection{The GRB host galaxy}
The abundance analysis shows that the GRB host absorber is a relatively
low-metallicity system (about 2\% solar metallicity) compared to the other GRB-DLAs at $z>4.5$ (GRB~100219A and GRB~050904), see Fig. \ref{Comparison}. The relative abundances are close to solar, as shown in Table \ref{table:metallicity}, but we are unable to address whether the metals are depleted onto dust grains, since we have no reliable measurement of $[$Zn$/$Fe$]$ \citep[typically decreasing with metallicity in both QSO- and GRB-DLAs:][]{2005ARA&A..43..861W, 2003ApJ...585..638S, Dessauges06, 2007ApJ...666..267P, 2008A&A...481..327N,2012ApJ...755...89R}, which is normally used to determine the amount of dust depletion.

In Fig.~\ref{RafelskiComparison} the two high-$z$ GRBs 100219A and 111008A (i.e. the only two GRBs with $z>4.7$ and measured metallicities with errors smaller than 0.5 dex) are compared to the recently measured QSO-DLA metallicities (from \citealt{2013arXiv1310.6042R}) at similar redshifts. The GRB metallicities fall within the range spanned by QSO-DLAs, but typically in the upper end of the distribution. This trend is also present at lower redshift \citep{2008ApJ...683..321F}. This phenomenon can be explained by the fact that GRB afterglows generally probe the star-forming `hearts' of galaxies, while QSO-DLAs have a higher chance to probe the (less metal rich) outskirts of galaxies (see e.g. \citealt{2007ApJ...666..267P} and \citealt{2013arXiv1308.3249L}). It is remarkable, however, that the metallicity of GRB-DLAs does not seem to be as redshift dependent as the metallicity of QSO-DLAs. At lower redshifts the \ion{H}{1} column density of GRB absorbers is much larger than that of QSO-DLAs. 

Concerning $\alpha$-element
overabundance, \citet{2012ApJ...755...89R} argue that the
$\alpha$-to-iron-group abundance ratios in QSO-DLAs are consistent with those
of halo-stars in the Milky Way.  Unfortunately, with the lack of a reliably determined column density of zinc we cannot distinguish the scenario, where dust-depletion and $\alpha$-element overabundance both are present, from the scenario where $\alpha$-element overabundance and dust-depletion both are absent.

address the issue of $\alpha$-elements overabundance.

The afterglow spectrum of the high-redshift GRB\,100219A ($z=4.7$) studied by \citet{2013MNRAS.428.3590T} reveals a substantially higher metallicity of [M/H] = $-1.0\pm0.1$ and evidence for either depletion on dust grains or a strong $\alpha$-element overabundance ([S/Fe] $= 0.8$). The GRB\,100219A absorber is also substantially more complex with five velocity
components spread over 160 km s$^{-1}$ whereas we only require two components
separated by 70 km s$^{-1}$ to fit the metal lines from the GRB\,111008A
absorber. Higher velocity widths for GRB\,100219A compared to GRB\,111008A are also expected due to the relation between velocity width and metallicity \citep{2006A&A...457...71L,2013ApJ...769...54N}.

The study by \citet{2013arXiv1310.6042R} measures metallicities of 17 QSO-DLAs with $z=3.64-5.08$. They show that the metallicity of QSO-DLAs decreases rapidly at $z>4.7$, likely because the state of the gas in the outer part of DLA-galaxies changes at this redshift. It is unknown whether GRB-DLAs exhibit the same trend. So far the only GRB-DLAs with a well-determined metallicity at $z>4.7$ are GRB 111008A and GRB 100219A. Since GRB-DLA sight lines have a different origin than QSO-DLAs it is not evident whether or not a similar decline in the metallicity of GRB-DLAs is be present.

\subsection{The intervening DLA}

The $z=4.6$ intervening DLA absorber has a high neutral hydrogen column density, which is consistent with being among the highest column densities of the 18 QSO-DLAs observed at similar high redshifts by \citet{2013arXiv1310.6042R}. A high metallicity is also indicated by the large equivalent width of \ion{Si}{2} $\lambda$1526. This suggests that this is not a typical intervening system likely associated with the outer regions of the absorber \citep{2007ApJ...666..267P}. It is possible that the line of sight towards the GRB crossed inner parts of the foreground absorber, thus showing higher column densities and metallicity. Further observations can potentially reveal the nature of the intervening absorber.

\section{Summary}

With spectroscopy of the GRB\,111008A  afterglow we have a rare chance of studying the properties of a sight line originating in a star forming region of a $z=5.0$ galaxy. By analyzing absorption lines from the interstellar medium of the GRB's host galaxy a metallicity of $[$S$/$H$]= -1.70 \pm 0.10$ is measured, and from fitting the spectral energy distribution the dust extinction is determined to be $A_V = 0.11 \pm 0.04$ mag. The dust-to-metals ratio is equal to or lower than what is observed in the Local Group. Determination of the dust-to-metals ratio of such high-redshift environments is important, since it can potentially constrain dust production mechanisms. 

GRB\,111008A offers two noteworthy features: it is the highest redshift GRB host galaxy with such a precise metallicity measurement, and it is also the first time fine structure lines from \ion{Fe}{2} have been observed at such a high redshift. This is special because the signal-to-noise required to detect the relatively weak fine-structure lines is often not reached at high redshift. Their presence unambiguously confirms the absorption system as the GRB's host galaxy.

In the sight line towards the GRB is also a DLA at $z=4.6$. The metallicity of this system is constrained by [Si/H]$>-2.15$ at a 2$\sigma$ limit, [Fe/H]$=-1.61\pm 0.17$ and [Ni/H] $=-1.33\pm 0.12$. The role of dust depletion is unclear. With photometric observations of the field surrounding the GRB's position performed roughly two years after explosion we detect emission from a source, which could be the intervening system. The offset between the detected source and the GRB sight line would be $4\pm 1$ kpc at the redshift of the intervening DLA. Deeper observations could potentially reveal whether or not this source is related to the intervening DLA at $z=4.6$.

\section*{Acknowledgement}

We thank the referee for very useful comments. We would also like to thank A.~De Cia for useful comments and discussions. The Dark Cosmology Centre is funded by the DNRF. This research is partly based on observations obtained at the Gemini Observatory, which is operated by the Association Research in Astronomy, Inc., under a cooperative agreement with the NSF on behalf of the Gemini partnership. OEH acknowledges the Dutch Research School for Astronomy (NOVA) for a PhD grant.
TK acknowledges support by the European Commission
under the Marie Curie Intra-European Fellowship Programme
in FP7. 
JPUF and BMJ acknowledge support from the ERC-StG grant EGGS-278202. Part of the funding for GROND (both hardware as well as personnel) was generously granted from the Leibniz-Prize to Prof. G. Hasinger (DFG grant HA 1850/28-1). SK acknowledges support by DFG grant Kl 766/16-1, and MN by DFG grant SA 2001/2-1. SS acknowledges support from the Iniciativa Cientifica Milenio grant P10-064-F (Millennium Center for Supernova Science), with input from ``Fondo de Innovaci\'{o}n para la Competitividad, del Ministerio de Econom\'{\i}a, Fomento y Turismo de Chile", and Basal-CATA (PFB-06/2007).

\bibliographystyle{apj}

\begin{thebibliography}{96}
\expandafter\ifx\csname natexlab\endcsname\relax\def\natexlab#1{#1}\fi

\bibitem[{{Aihara} {et~al.}(2011){Aihara}, {Allende Prieto}, {An}, {Anderson},
  {Aubourg}, {Balbinot}, {Beers}, {Berlind}, {Bickerton}, {Bizyaev}, {Blanton},
  {Bochanski}, {Bolton}, {Bovy}, {Brandt}, {Brinkmann}, {Brown}, {Brownstein},
  {Busca}, {Campbell}, {Carr}, {Chen}, {Chiappini}, {Comparat}, {Connolly},
  {Cortes}, {Croft}, {Cuesta}, {da Costa}, {Davenport}, {Dawson}, {Dhital},
  {Ealet}, {Ebelke}, {Edmondson}, {Eisenstein}, {Escoffier}, {Esposito},
  {Evans}, {Fan}, {Femen{\'{\i}}a Castell{\'a}}, {Font-Ribera}, {Frinchaboy},
  {Ge}, {Gillespie}, {Gilmore}, {Gonz{\'a}lez Hern{\'a}ndez}, {Gott}, {Gould},
  {Grebel}, {Gunn}, {Hamilton}, {Harding}, {Harris}, {Hawley}, {Hearty}, {Ho},
  {Hogg}, {Holtzman}, {Honscheid}, {Inada}, {Ivans}, {Jiang}, {Johnson},
  {Jordan}, {Jordan}, {Kazin}, {Kirkby}, {Klaene}, {Knapp}, {Kneib},
  {Kochanek}, {Koesterke}, {Kollmeier}, {Kron}, {Lampeitl}, {Lang}, {Le Goff},
  {Lee}, {Lin}, {Long}, {Loomis}, {Lucatello}, {Lundgren}, {Lupton}, {Ma},
  {MacDonald}, {Mahadevan}, {Maia}, {Makler}, {Malanushenko}, {Malanushenko},
  {Mandelbaum}, {Maraston}, {Margala}, {Masters}, {McBride}, {McGehee},
  {McGreer}, {M{\'e}nard}, {Miralda-Escud{\'e}}, {Morrison}, {Mullally},
  {Muna}, {Munn}, {Murayama}, {Myers}, {Naugle}, {Neto}, {Nguyen}, {Nichol},
  {O'Connell}, {Ogando}, {Olmstead}, {Oravetz}, {Padmanabhan},
  {Palanque-Delabrouille}, {Pan}, {Pandey}, {P{\^a}ris}, {Percival},
  {Petitjean}, {Pfaffenberger}, {Pforr}, {Phleps}, {Pichon}, {Pieri}, {Prada},
  {Price-Whelan}, {Raddick}, {Ramos}, {Reyl{\'e}}, {Rich}, {Richards}, {Rix},
  {Robin}, {Rocha-Pinto}, {Rockosi}, {Roe}, {Rollinde}, {Ross}, {Ross},
  {Rossetto}, {S{\'a}nchez}, {Sayres}, {Schlegel}, {Schlesinger}, {Schmidt},
  {Schneider}, {Sheldon}, {Shu}, {Simmerer}, {Simmons}, {Sivarani}, {Snedden},
  {Sobeck}, {Steinmetz}, {Strauss}, {Szalay}, {Tanaka}, {Thakar}, {Thomas},
  {Tinker}, {Tofflemire}, {Tojeiro}, {Tremonti}, {Vandenberg}, {Vargas
  Maga{\~n}a}, {Verde}, {Vogt}, {Wake}, {Wang}, {Weaver}, {Weinberg}, {White},
  {White}, {Yanny}, {Yasuda}, {Yeche}, \& {Zehavi}}]{2011ApJS..193...29A}
{Aihara}, H., {et~al.} 2011, \apjs, 193, 29

\bibitem[{{Appenzeller} {et~al.}(1998){Appenzeller}, {Fricke}, {F{\"u}rtig},
  {G{\"a}ssler}, {H{\"a}fner}, {Harke}, {Hess}, {Hummel}, {J{\"u}rgens},
  {Kudritzki}, {Mantel}, {Meisl}, {Muschielok}, {Nicklas}, {Rupprecht},
  {Seifert}, {Stahl}, {Szeifert}, \& {Tarantik}}]{1998Msngr..94....1A}
{Appenzeller}, I., {et~al.} 1998, The Messenger, 94, 1

\bibitem[{{Asplund} {et~al.}(2009){Asplund}, {Grevesse}, {Sauval}, \&
  {Scott}}]{2009ARA&A..47..481A}
{Asplund}, M., {Grevesse}, N., {Sauval}, A.~J., \& {Scott}, P. 2009, \araa, 47,
  481

\bibitem[{{Basa} {et~al.}(2012){Basa}, {Cuby}, {Savaglio}, {Boissier},
  {Cl{\'e}ment}, {Flores}, {Le Borgne}, \& {Mazure}}]{2012A&A...542A.103B}
{Basa}, S., {Cuby}, J.~G., {Savaglio}, S., {Boissier}, S., {Cl{\'e}ment}, B.,
  {Flores}, H., {Le Borgne}, D., \& {Mazure}, A. 2012, \aap, 542, A103

\bibitem[{{Baumgartner} {et~al.}(2011){Baumgartner}, {Barthelmy}, {Cummings},
  {Fenimore}, {Gehrels}, {Krimm}, {Markwardt}, {Palmer}, {Sakamoto}, {Saxton},
  {Stamatikos}, {Tueller}, \& {Ukwatta}}]{2011GCN..12424...1B}
{Baumgartner}, W.~H., {et~al.} 2011, GRB Coordinates Network, 12424, 1

\bibitem[{{Beardmore} {et~al.}(2011){Beardmore}, {Evans}, {Goad}, \&
  {Osborne}}]{2011GCN..12425...1B}
{Beardmore}, A.~P., {Evans}, P.~A., {Goad}, M.~R., \& {Osborne}, J.~P. 2011,
  GRB Coordinates Network, 12425, 1

\bibitem[{{Berger} {et~al.}(2011){Berger}, {Chornock}, {Holmes}, {Foley},
  {Cucchiara}, {Wolf}, {Podsiadlowski}, {Fox}, \& {Roth}}]{2011ApJ...743..204B}
{Berger}, E., {et~al.} 2011, \apj, 743, 204

\bibitem[{{Berger} {et~al.}(2006){Berger}, {Penprase}, {Cenko}, {Kulkarni},
  {Fox}, {Steidel}, \& {Reddy}}]{2006ApJ...642..979B}
{Berger}, E., {Penprase}, B.~E., {Cenko}, S.~B., {Kulkarni}, S.~R., {Fox},
  D.~B., {Steidel}, C.~C., \& {Reddy}, N.~A. 2006, \apj, 642, 979

\bibitem[{{Bloom} {et~al.}(2002){Bloom}, {Kulkarni}, \&
  {Djorgovski}}]{2002AJ....123.1111B}
{Bloom}, J.~S., {Kulkarni}, S.~R., \& {Djorgovski}, S.~G. 2002, \aj, 123, 1111

\bibitem[{{Bouwens} {et~al.}(2007){Bouwens}, {Illingworth}, {Franx}, \&
  {Ford}}]{2007ApJ...670..928B}
{Bouwens}, R.~J., {Illingworth}, G.~D., {Franx}, M., \& {Ford}, H. 2007, \apj,
  670, 928

\bibitem[{{Campana} {et~al.}(2006){Campana}, {Mangano}, {Blustin}, {Brown},
  {Burrows}, {Chincarini}, {Cummings}, {Cusumano}, {Della Valle}, {Malesani},
  {M{\'e}sz{\'a}ros}, {et~al.}}]{2006Natur.442.1008C}
{Campana}, S., {et~al.} 2006, \nat, 442, 1008

\bibitem[{{Chary} {et~al.}(2007){Chary}, {Berger}, \&
  {Cowie}}]{2007ApJ...671..272C}
{Chary}, R., {Berger}, E., \& {Cowie}, L. 2007, \apj, 671, 272

\bibitem[{{Chen} {et~al.}(2013){Chen}, {Dai}, {Kochanek}, \&
  {Chartas}}]{2013arXiv1306.0008C}
{Chen}, B., {Dai}, X., {Kochanek}, C.~S., \& {Chartas}, G. 2013,
  astroph/1306.0008

\bibitem[{{Chornock} {et~al.}(2013){Chornock}, {Berger}, {Fox}, {Lunnan},
  {Drout}, {Fong}, {Laskar}, \& {Roth}}]{2013ApJ...774...26C}
{Chornock}, R., {Berger}, E., {Fox}, D.~B., {Lunnan}, R., {Drout}, M.~R.,
  {Fong}, W.-f., {Laskar}, T., \& {Roth}, K.~C. 2013, \apj, 774, 26

\bibitem[{{Covino} {et~al.}(2013){Covino}, {Melandri}, {Salvaterra}, {Campana},
  {Vergani}, {Bernardini}, {D'Avanzo}, {D'Elia}, {Fugazza}, {Ghirlanda},
  {Ghisellini}, {Gomboc}, {Jin}, {Kr{\"u}hler}, {Malesani}, {Nava},
  {Sbarufatti}, \& {Tagliaferri}}]{2013MNRAS.432.1231C}
{Covino}, S., {et~al.} 2013, \mnras, 432, 1231

\bibitem[{{Cucchiara} {et~al.}(2011){Cucchiara}, {Levan}, {Fox}, {Tanvir},
  {Ukwatta}, {Berger}, {Kr{\"u}hler}, {K{\"u}pc{\"u} Yolda{\c s}}, {Wu},
  {Toma}, {Greiner}, {Olivares}, {Rowlinson}, {Amati}, {Sakamoto}, {Roth},
  {Stephens}, {Fritz}, {Fynbo}, {Hjorth}, {Malesani}, {Jakobsson}, {Wiersema},
  {O'Brien}, {Soderberg}, {Foley}, {Fruchter}, {Rhoads}, {Rutledge}, {Schmidt},
  {Dopita}, {Podsiadlowski}, {Willingale}, {Wolf}, {Kulkarni}, \&
  {D'Avanzo}}]{2011ApJ...736....7C}
{Cucchiara}, A., {et~al.} 2011, \apj, 736, 7

\bibitem[{{D'Avanzo} {et~al.}(2010){D'Avanzo}, {Perri}, {Fugazza},
  {Salvaterra}, {Chincarini}, {Margutti}, {Wu}, {Th{\"o}ne},
  {Fern{\'a}ndez-Soto}, {Ukwatta}, {Burrows}, {Gehrels}, {Meszaros}, {Toma},
  {Zhang}, {Covino}, {Campana}, {D'Elia}, {Della Valle}, \&
  {Piranomonte}}]{2010A&A...522A..20D}
{D'Avanzo}, P., {et~al.} 2010, \aap, 522, A20

\bibitem[{{De Cia} {et~al.}(2011){De Cia}, {Jakobsson}, {Bj{\"o}rnsson},
  {Vreeswijk}, {Dhillon}, {Marsh}, {Chapman}, {Fynbo}, {Ledoux}, {Littlefair},
  {Malesani}, {Schulze}, {Smette}, {Zafar}, \&
  {Gudmundsson}}]{2011MNRAS.412.2229D}
{De Cia}, A., {et~al.} 2011, \mnras, 412, 2229

\bibitem[{{D'Elia} {et~al.}(2012){D'Elia}, {Campana}, {Covino}, {D'Avanzo},
  {Piranomonte}, \& {Tagliaferri}}]{2012MSAIS..21..206D}
{D'Elia}, V., {Campana}, S., {Covino}, S., {D'Avanzo}, P., {Piranomonte}, S.,
  \& {Tagliaferri}, G. 2012, Memorie della Societa Astronomica Italiana
  Supplementi, 21, 206

\bibitem[{{D'Elia} {et~al.}(2010){D'Elia}, {Fynbo}, {Covino}, {Goldoni},
  {Jakobsson}, {Matteucci}, {Piranomonte}, {Sollerman}, {Th{\"o}ne}, {Vergani},
  {Vreeswijk}, {Watson}, {Wiersema}, {Zafar}, {de Ugarte Postigo}, {Flores},
  {Hjorth}, {Kaper}, {Levan}, {Malesani}, {Milvang-Jensen}, {Pian},
  {Tagliaferri}, \& {Tanvir}}]{2010A&A...523A..36D}
{D'Elia}, V., {et~al.} 2010, \aap, 523, A36

\bibitem[{{D'Elia} {et~al.}(2014){D'Elia}, {Fynbo}, {Goldoni}, {Covino}, {de
  Ugarte Postigo}, {Ledoux}, {Calura}, {Gorosabel}, {Malesani},
  {Sanchez-Ramirez}, {Savaglio}, {Castro-Tirado}, {Hartoog}, {Kaper},
  {Munoz-Darias}, {Pian}, {Piranomonte}, {Tagliaferri}, {Tanvir}, {Vergani},
  {Watson}, \& {Xu}}]{2014arXiv1402.4026D}
---. 2014, ArXiv astroph/1402.4026

\bibitem[{{Dessauges-Zavadsky} {et~al.}(2006){Dessauges-Zavadsky}, {Prochaska},
  {D'Odorico}, {Calura}, \& {Matteucci}}]{Dessauges06}
{Dessauges-Zavadsky}, M., {Prochaska}, J.~X., {D'Odorico}, S., {Calura}, F., \&
  {Matteucci}, F. 2006, \aap, 445, 93

\bibitem[{{Draine}(2009)}]{2009ASPC..414..453D}
{Draine}, B.~T. 2009, in Astronomical Society of the Pacific Conference Series,
  Vol. 414, Cosmic Dust - Near and Far, ed. T.~{Henning}, E.~{Gr{\"u}n}, \&
  J.~{Steinacker}, 453

\bibitem[{{Evans} {et~al.}(2009){Evans}, {Beardmore}, {Page}, {Osborne},
  {O'Brien}, {Willingale}, {Starling}, {Burrows}, {Godet}, {Vetere}, {Racusin},
  {Goad}, {Wiersema}, {Angelini}, {Capalbi}, {Chincarini}, {Gehrels}, {Kennea},
  {Margutti}, {Morris}, {Mountford}, {Pagani}, {Perri}, {Romano}, \&
  {Tanvir}}]{2009MNRAS.397.1177E}
{Evans}, P.~A., {et~al.} 2009, \mnras, 397, 1177

\bibitem[{{Evans} {et~al.}(2007){Evans}, {Beardmore}, {Page}, {Tyler},
  {Osborne}, {Goad}, {O'Brien}, {Vetere}, {Racusin}, {Morris}, {Burrows},
  {Capalbi}, {Perri}, {Gehrels}, \& {Romano}}]{2007A&A...469..379E}
---. 2007, \aap, 469, 379

\bibitem[{{Fiore} {et~al.}(2005){Fiore}, {D'Elia}, {Lazzati}, {Perna},
  {Sbordone}, {Stratta}, {Meurs}, {Ward}, {Antonelli}, {Chincarini}, {Covino},
  {Di Paola}, {Fontana}, {Ghisellini}, {Israel}, {Frontera}, {Marconi},
  {Stella}, {Vietri}, \& {Zerbi}}]{2005ApJ...624..853F}
{Fiore}, F., {et~al.} 2005, \apj, 624, 853

\bibitem[{{Foreman-Mackey} {et~al.}(2013){Foreman-Mackey}, {Hogg}, {Lang}, \&
  {Goodman}}]{2013PASP..125..306F}
{Foreman-Mackey}, D., {Hogg}, D.~W., {Lang}, D., \& {Goodman}, J. 2013, \pasp,
  125, 306

\bibitem[{{Fynbo} {et~al.}(2009){Fynbo}, {Jakobsson}, {Prochaska}, {Malesani},
  {Ledoux}, {de Ugarte Postigo}, {Nardini}, {Vreeswijk}, {Wiersema}, {Hjorth},
  {Sollerman}, {Chen}, {Th{\"o}ne}, {Bj{\"o}rnsson}, {Bloom}, {Castro-Tirado},
  {Christensen}, {De Cia}, {Fruchter}, {Gorosabel}, {Graham}, {Jaunsen},
  {Jensen}, {Kann}, {Kouveliotou}, {Levan}, {Maund}, {Masetti},
  {Milvang-Jensen}, {Palazzi}, {Perley}, {Pian}, {Rol}, {Schady}, {Starling},
  {Tanvir}, {Watson}, {Xu}, {Augusteijn}, {Grundahl}, {Telting}, \&
  {Quirion}}]{2009ApJS..185..526F}
{Fynbo}, J.~P.~U., {et~al.} 2009, \apjs, 185, 526

\bibitem[{{Fynbo} {et~al.}(2008){Fynbo}, {Prochaska}, {Sommer-Larsen},
  {Dessauges-Zavadsky}, \& {M{\o}ller}}]{2008ApJ...683..321F}
{Fynbo}, J.~P.~U., {Prochaska}, J.~X., {Sommer-Larsen}, J.,
  {Dessauges-Zavadsky}, M., \& {M{\o}ller}, P. 2008, \apj, 683, 321

\bibitem[{{Fynbo} {et~al.}(2006){Fynbo}, {Starling}, {Ledoux}, {Wiersema},
  {Th{\"o}ne}, {Sollerman}, {Jakobsson}, {Hjorth}, {Watson}, {Vreeswijk},
  {M{\o}ller}, {Rol}, {Gorosabel}, {N{\"a}r{\"a}nen}, {Wijers},
  {Bj{\"o}rnsson}, {Castro Cer{\'o}n}, {Curran}, {Hartmann}, {Holland},
  {Jensen}, {Levan}, {Limousin}, {Kouveliotou}, {Nelemans}, {Pedersen},
  {Priddey}, \& {Tanvir}}]{2006A&A...451L..47F}
{Fynbo}, J.~P.~U., {et~al.} 2006, \aap, 451, L47

\bibitem[{{Galama} {et~al.}(1998){Galama}, {Vreeswijk}, {van Paradijs},
  {Kouveliotou}, {Augusteijn}, {B{\"o}hnhardt}, {Brewer}, {Doublier},
  {Gonzalez}, {Leibundgut}, {Lidman}, {Hainaut}, {Patat}, {Heise}, {in't Zand},
  {Hurley}, {Groot}, {Strom}, {Mazzali}, {Iwamoto}, {Nomoto}, {Umeda},
  {Nakamura}, {Young}, {Suzuki}, {Shigeyama}, {Koshut}, {Kippen}, {Robinson},
  {de Wildt}, {Wijers}, {Tanvir}, {Greiner}, {Pian}, {Palazzi}, {Frontera},
  {Masetti}, {Nicastro}, {Feroci}, {Costa}, {Piro}, {Peterson}, {Tinney},
  {Boyle}, {Cannon}, {Stathakis}, {Sadler}, {Begam}, \&
  {Ianna}}]{1998Natur.395..670G}
{Galama}, T.~J., {et~al.} 1998, \nat, 395, 670

\bibitem[{{Greiner} {et~al.}(2008){Greiner}, {Bornemann}, {Clemens}, {Deuter},
  {Hasinger}, {Honsberg}, {Huber}, {Huber}, {Krauss}, {Kr{\"u}hler},
  {K{\"u}pc{\"u} Yolda{\c s}}, {Mayer-Hasselwander}, {Mican}, {Primak},
  {Schrey}, {Steiner}, {Szokoly}, {Th{\"o}ne}, {Yolda{\c s}}, {Klose}, {Laux},
  \& {Winkler}}]{2008PASP..120..405G}
{Greiner}, J., {et~al.} 2008, \pasp, 120, 405

\bibitem[{{Greiner} {et~al.}(2011){Greiner}, {Kr{\"u}hler}, {Klose}, {Afonso},
  {Clemens}, {Filgas}, {Hartmann}, {K{\"u}pc{\"u} Yolda{\c s}}, {Nardini},
  {Olivares E.}, {Rau}, {Rossi}, {Schady}, \& {Updike}}]{2011A&A...526A..30G}
---. 2011, \aap, 526, A30

\bibitem[{{Hartoog} {et~al.}(2013){Hartoog}, {Wiersema}, {Vreeswijk}, {Kaper},
  {Tanvir}, {Savaglio}, {Berger}, {Chornock}, {Covino}, {D'Elia}, {Flores},
  {Fynbo}, {Goldoni}, {Gomboc}, {Melandri}, {Pozanenko}, {Schaye}, {Postigo},
  \& {Wijers}}]{hartoog2013}
{Hartoog}, O.~E., {et~al.} 2013, \mnras, 430, 2739

\bibitem[{{Herrera-Camus} {et~al.}(2012){Herrera-Camus}, {Fisher}, {Bolatto},
  {Leroy}, {Walter}, {Gordon}, {Roman-Duval}, {Donaldson}, {Mel{\'e}ndez}, \&
  {Cannon}}]{2012ApJ...752..112H}
{Herrera-Camus}, R., {et~al.} 2012, \apj, 752, 112

\bibitem[{{Hjorth} \& {Bloom}(2012)}]{2012grbu.book..169H}
{Hjorth}, J., \& {Bloom}, J.~S. 2012, {The Gamma-Ray Burst - Supernova
  Connection} (Cambridge University Press, Cambridge), 169--190

\bibitem[{{Hjorth} {et~al.}(2012){Hjorth}, {Malesani}, {Jakobsson}, {Jaunsen},
  {Fynbo}, {Gorosabel}, {Kr{\"u}hler}, {Levan}, {Micha{\l}owski},
  {Milvang-Jensen}, {M{\o}ller}, {Schulze}, {Tanvir}, \&
  {Watson}}]{2012ApJ...756..187H}
{Hjorth}, J., {et~al.} 2012, \apj, 756, 187

\bibitem[{{Hjorth} {et~al.}(2003){Hjorth}, {Sollerman}, {M{\o}ller}, {Fynbo},
  {Woosley}, {Kouveliotou}, {Tanvir}, {Greiner}, {Andersen}, {Castro-Tirado},
  {Castro Cer{\'o}n}, {Fruchter}, {Gorosabel}, {Jakobsson}, {Kaper}, {Klose},
  {Masetti}, {Pedersen}, {Pedersen}, {Pian}, {Palazzi}, {Rhoads}, {Rol}, {van
  den Heuvel}, {Vreeswijk}, {Watson}, \& {Wijers}}]{2003Natur.423..847H}
---. 2003, \nat, 423, 847

\bibitem[{{Horne}(1986)}]{1986PASP...98..609H}
{Horne}, K. 1986, \pasp, 98, 609

\bibitem[{{Jakobsson} {et~al.}(2004){Jakobsson}, {Hjorth}, {Fynbo}, {Watson},
  {Pedersen}, {Bj{\"o}rnsson}, \& {Gorosabel}}]{2004ApJ...617L..21J}
{Jakobsson}, P., {Hjorth}, J., {Fynbo}, J.~P.~U., {Watson}, D., {Pedersen}, K.,
  {Bj{\"o}rnsson}, G., \& {Gorosabel}, J. 2004, \apjl, 617, L21

\bibitem[{{Kalberla} {et~al.}(2005){Kalberla}, {Burton}, {Hartmann}, {Arnal},
  {Bajaja}, {Morras}, \& {P{\"o}ppel}}]{2005A&A...440..775K}
{Kalberla}, P.~M.~W., {Burton}, W.~B., {Hartmann}, D., {Arnal}, E.~M.,
  {Bajaja}, E., {Morras}, R., \& {P{\"o}ppel}, W.~G.~L. 2005, \aap, 440, 775

\bibitem[{{Kawai} {et~al.}(2006){Kawai}, {Kosugi}, {Aoki}, {Yamada}, {Totani},
  {Ohta}, {Iye}, {Hattori}, {Aoki}, {Furusawa}, {Hurley}, {Kawabata},
  {Kobayashi}, {Komiyama}, {Mizumoto}, {Nomoto}, {Noumaru}, {Ogasawara},
  {Sato}, {Sekiguchi}, {Shirasaki}, {Suzuki}, {Takata}, {Tamagawa}, {Terada},
  {Watanabe}, {Yatsu}, \& {Yoshida}}]{2006Natur.440..184K}
{Kawai}, N., {et~al.} 2006, \nat, 440, 184

\bibitem[{{Kr{\"u}hler} {et~al.}(2011){Kr{\"u}hler}, {Greiner}, {Schady},
  {Savaglio}, {Afonso}, {Clemens}, {Elliott}, {Filgas}, {Gruber}, {Kann},
  {Klose}, {K{\"u}pc{\"u}-Yolda{\c s}}, {McBreen}, {Olivares}, {Pierini},
  {Rau}, {Rossi}, {Nardini}, {Nicuesa Guelbenzu}, {Sudilovsky}, \&
  {Updike}}]{2011A&A...534A.108K}
{Kr{\"u}hler}, T., {et~al.} 2011, \aap, 534, A108

\bibitem[{{Kr{\"u}hler} {et~al.}(2008){Kr{\"u}hler}, {K{\"u}pc{\"u} Yolda{\c
  s}}, {Greiner}, {Clemens}, {McBreen}, {Primak}, {Savaglio}, {Yolda{\c s}},
  {Szokoly}, \& {Klose}}]{2008ApJ...685..376K}
---. 2008, \apj, 685, 376

\bibitem[{{Kr{\"u}hler} {et~al.}(2013){Kr{\"u}hler}, {Ledoux}, {Fynbo},
  {Vreeswijk}, {Schmidl}, {Malesani}, {Christensen}, {De Cia}, {Hjorth},
  {Jakobsson}, {Kann}, {Kaper}, {Vergani}, {Afonso}, {Covino}, {de Ugarte
  Postigo}, {D'Elia}, {Filgas}, {Goldoni}, {Greiner}, {Hartoog},
  {Milvang-Jensen}, {Nardini}, {Piranomonte}, {Rossi},
  {S{\'a}nchez-Ram{\'{\i}}rez}, {Schady}, {Schulze}, {Sudilovsky}, {Tanvir},
  {Tagliaferri}, {Watson}, {Wiersema}, {Wijers}, \& {Xu}}]{2013A&A...557A..18K}
---. 2013, \aap, 557, A18

\bibitem[{{Ledoux} {et~al.}(2006){Ledoux}, {Petitjean}, {Fynbo}, {M{\o}ller},
  \& {Srianand}}]{2006A&A...457...71L}
{Ledoux}, C., {Petitjean}, P., {Fynbo}, J.~P.~U., {M{\o}ller}, P., \&
  {Srianand}, R. 2006, \aap, 457, 71

\bibitem[{{Ledoux} {et~al.}(2009){Ledoux}, {Vreeswijk}, {Smette}, {Fox},
  {Petitjean}, {Ellison}, {Fynbo}, \& {Savaglio}}]{2009A&A...506..661L}
{Ledoux}, C., {Vreeswijk}, P.~M., {Smette}, A., {Fox}, A.~J., {Petitjean}, P.,
  {Ellison}, S.~L., {Fynbo}, J.~P.~U., \& {Savaglio}, S. 2009, \aap, 506, 661

\bibitem[{{Lemasle} {et~al.}(2013){Lemasle}, {Fran{\c c}ois}, {Genovali},
  {Kovtyukh}, {Bono}, {Inno}, {Laney}, {Kaper}, {Bergemann}, {Fabrizio},
  {Matsunaga}, {Pedicelli}, {Primas}, \& {Romaniello}}]{2013arXiv1308.3249L}
{Lemasle}, B., {et~al.} 2013, \aap, 558, A31

\bibitem[{{Levan} {et~al.}(2011{\natexlab{a}}){Levan}, {Wiersema}, \&
  {Tanvir}}]{2011GCN..12426...1L}
{Levan}, A.~J., {Wiersema}, K., \& {Tanvir}, N.~R. 2011{\natexlab{a}}, GRB
  Coordinates Network, 12426, 1

\bibitem[{{Levan} {et~al.}(2011{\natexlab{b}}){Levan}, {Wiersema}, \&
  {Tanvir}}]{2011GCN..12429...1L}
---. 2011{\natexlab{b}}, GRB Coordinates Network, 12429, 1

\bibitem[{{Matteucci} \& {Greggio}(1986)}]{1986A&A...154..279M}
{Matteucci}, F., \& {Greggio}, L. 1986, \aap, 154, 279

\bibitem[{{Mattsson} {et~al.}(2012){Mattsson}, {Andersen}, \&
  {Munkhammar}}]{Mattsson12}
{Mattsson}, L., {Andersen}, A.~C., \& {Munkhammar}, J.~D. 2012, \mnras, 423, 26

\bibitem[{{Modigliani} {et~al.}(2010){Modigliani}, {Goldoni}, {Royer},
  {Haigron}, {Guglielmi}, {Fran{\c c}ois}, {Horrobin}, {Bristow}, {Vernet},
  {Moehler}, {Kerber}, {Ballester}, {Mason}, \&
  {Christensen}}]{2010SPIE.7737E..56M}
{Modigliani}, A., {et~al.} 2010, in Society of Photo-Optical Instrumentation
  Engineers (SPIE) Conference Series, Vol. 7737, Society of Photo-Optical
  Instrumentation Engineers (SPIE) Conference Series

\bibitem[{{Modjaz} {et~al.}(2006){Modjaz}, {Stanek}, {Garnavich}, {Berlind},
  {Blondin}, {Brown}, {Calkins}, {Challis}, {Diamond-Stanic},
  {et~al.}}]{2006ApJ...645L..21M}
{Modjaz}, M., {et~al.} 2006, \apjl, 645, L21

\bibitem[{{Morgan} \& {Edmunds}(2003)}]{2003MNRAS.343..427M}
{Morgan}, H.~L., \& {Edmunds}, M.~G. 2003, \mnras, 343, 427

\bibitem[{{Nardini} {et~al.}(2011){Nardini}, {Klose}, {Greiner}, \&
  {Afonso}}]{2011GCN..12428...1N}
{Nardini}, M., {Klose}, S., {Greiner}, J., \& {Afonso}, P. 2011, GRB
  Coordinates Network, 12428, 1

\bibitem[{{Neeleman} {et~al.}(2013){Neeleman}, {Wolfe}, {Prochaska}, \&
  {Rafelski}}]{2013ApJ...769...54N}
{Neeleman}, M., {Wolfe}, A.~M., {Prochaska}, J.~X., \& {Rafelski}, M. 2013,
  \apj, 769, 54

\bibitem[{{Noterdaeme} {et~al.}(2008){Noterdaeme}, {Ledoux}, {Petitjean}, \&
  {Srianand}}]{2008A&A...481..327N}
{Noterdaeme}, P., {Ledoux}, C., {Petitjean}, P., \& {Srianand}, R. 2008, \aap,
  481, 327

\bibitem[{{Pei}(1992)}]{1992ApJ...395..130P}
{Pei}, Y.~C. 1992, \apj, 395, 130

\bibitem[{{Perley} {et~al.}(2009){Perley}, {Cenko}, {Bloom}, {Chen}, {Butler},
  {Kocevski}, {Prochaska}, {Brodwin}, {Glazebrook}, {Kasliwal}, {Kulkarni},
  {Lopez}, {Ofek}, {Pettini}, {Soderberg}, \& {Starr}}]{2009AJ....138.1690P}
{Perley}, D.~A., {et~al.} 2009, \aj, 138, 1690

\bibitem[{{Price} {et~al.}(2007){Price}, {Songaila}, {Cowie}, {Bell Burnell},
  {Berger}, {Cucchiara}, {Fox}, {Hook}, {Kulkarni}, {Penprase}, {Roth}, \&
  {Schmidt}}]{2007ApJ...663L..57P}
{Price}, P.~A., {et~al.} 2007, \apjl, 663, L57

\bibitem[{{Prochaska} {et~al.}(2006){Prochaska}, {Chen}, \&
  {Bloom}}]{prochaska2006}
{Prochaska}, J.~X., {Chen}, H.-W., \& {Bloom}, J.~S. 2006, \apj, 648, 95

\bibitem[{{Prochaska} {et~al.}(2007{\natexlab{a}}){Prochaska}, {Chen}, {Bloom},
  {Dessauges-Zavadsky}, {O'Meara}, {Foley}, {Bernstein}, {Burles}, {Dupree},
  {Falco}, \& {Thompson}}]{2007ApJS..168..231P}
{Prochaska}, J.~X., {et~al.} 2007{\natexlab{a}}, \apjs, 168, 231

\bibitem[{{Prochaska} {et~al.}(2007{\natexlab{b}}){Prochaska}, {Chen},
  {Dessauges-Zavadsky}, \& {Bloom}}]{2007ApJ...666..267P}
{Prochaska}, J.~X., {Chen}, H.-W., {Dessauges-Zavadsky}, M., \& {Bloom}, J.~S.
  2007{\natexlab{b}}, \apj, 666, 267

\bibitem[{{Prochaska} {et~al.}(2008){Prochaska}, {Chen}, {Wolfe},
  {Dessauges-Zavadsky}, \& {Bloom}}]{2008ApJ...672...59P}
{Prochaska}, J.~X., {Chen}, H.-W., {Wolfe}, A.~M., {Dessauges-Zavadsky}, M., \&
  {Bloom}, J.~S. 2008, \apj, 672, 59

\bibitem[{{Prochaska} {et~al.}(2009){Prochaska}, {Sheffer}, {Perley}, {Bloom},
  {Lopez}, {Dessauges-Zavadsky}, {Chen}, {Filippenko}, {Ganeshalingam}, {Li},
  {Miller}, \& {Starr}}]{2009ApJ...691L..27P}
{Prochaska}, J.~X., {et~al.} 2009, \apjl, 691, L27

\bibitem[{{Rafelski} {et~al.}(2014){Rafelski}, {Neeleman}, {Fumagalli},
  {Wolfe}, \& {Prochaska}}]{2013arXiv1310.6042R}
{Rafelski}, M., {Neeleman}, M., {Fumagalli}, M., {Wolfe}, A.~M., \&
  {Prochaska}, J.~X. 2014, \apjl, 782, L29

\bibitem[{{Rafelski} {et~al.}(2012){Rafelski}, {Wolfe}, {Prochaska},
  {Neeleman}, \& {Mendez}}]{2012ApJ...755...89R}
{Rafelski}, M., {Wolfe}, A.~M., {Prochaska}, J.~X., {Neeleman}, M., \&
  {Mendez}, A.~J. 2012, \apj, 755, 89

\bibitem[{{Salvaterra} {et~al.}(2009){Salvaterra}, {Della Valle}, {Campana},
  {Chincarini}, {Covino}, {D'Avanzo}, {Fern{\'a}ndez-Soto}, {Guidorzi},
  {Mannucci}, {Margutti}, {Th{\"o}ne}, {Antonelli}, {Barthelmy}, {de Pasquale},
  {D'Elia}, {Fiore}, {Fugazza}, {Hunt}, {Maiorano}, {Marinoni}, {Marshall},
  {Molinari}, {Nousek}, {Pian}, {Racusin}, {Stella}, {Amati}, {Andreuzzi},
  {Cusumano}, {Fenimore}, {Ferrero}, {Giommi}, {Guetta}, {Holland}, {Hurley},
  {Israel}, {Mao}, {Markwardt}, {Masetti}, {Pagani}, {Palazzi}, {Palmer},
  {Piranomonte}, {Tagliaferri}, \& {Testa}}]{2009Natur.461.1258S}
{Salvaterra}, R., {et~al.} 2009, \nat, 461, 1258

\bibitem[{{Sari} {et~al.}(1998){Sari}, {Piran}, \&
  {Narayan}}]{1998ApJ...497L..17S}
{Sari}, R., {Piran}, T., \& {Narayan}, R. 1998, \apjl, 497, L17

\bibitem[{{Savaglio}(2006)}]{2006NJPh....8..195S}
{Savaglio}, S. 2006, New Journal of Physics, 8, 195

\bibitem[{{Savaglio} {et~al.}(2003){Savaglio}, {Fall}, \&
  {Fiore}}]{2003ApJ...585..638S}
{Savaglio}, S., {Fall}, S.~M., \& {Fiore}, F. 2003, \apj, 585, 638

\bibitem[{{Saxton} {et~al.}(2011){Saxton}, {Barthelmy}, {Beardmore}, {Holland},
  {Kennea}, {Krimm}, {Kuin}, {Littlejohns}, {Marshall}, {Pagani}, {Page},
  {Palmer}, \& {Swenson}}]{2011GCN..12423...1S}
{Saxton}, C.~J., {et~al.} 2011, GRB Coordinates Network, 12423, 1

\bibitem[{{Schady} {et~al.}(2011){Schady}, {Savaglio}, {Kr{\"u}hler},
  {Greiner}, \& {Rau}}]{2011A&A...525A.113S}
{Schady}, P., {Savaglio}, S., {Kr{\"u}hler}, T., {Greiner}, J., \& {Rau}, A.
  2011, \aap, 525, A113

\bibitem[{{Schulze} {et~al.}(2012){Schulze}, {Fynbo}, {Milvang-Jensen},
  {Rossi}, {Jakobsson}, {Ledoux}, {De Cia}, {Kr{\"u}hler}, {Mehner},
  {Bj{\"o}rnsson}, {Chen}, {Vreeswijk}, {Perley}, {Hjorth}, {Levan}, {Tanvir},
  {Ellison}, {M{\o}ller}, {Worseck}, {Chapman}, {Dall'Aglio}, \&
  {Letawe}}]{2012A&A...546A..20S}
{Schulze}, S., {et~al.} 2012, \aap, 546, A20

\bibitem[{{Skrutskie} {et~al.}(2006){Skrutskie}, {Cutri}, {Stiening},
  {Weinberg}, {Schneider}, {Carpenter}, {Beichman}, {Capps}, {Chester},
  {Elias}, {Huchra}, {Liebert}, {Lonsdale}, {Monet}, {Price}, {Seitzer},
  {Jarrett}, {Kirkpatrick}, {Gizis}, {Howard}, {Evans}, {Fowler}, {Fullmer},
  {Hurt}, {Light}, {Kopan}, {Marsh}, {McCallon}, {Tam}, {Van Dyk}, \&
  {Wheelock}}]{2006AJ....131.1163S}
{Skrutskie}, M.~F., {et~al.} 2006, \aj, 131, 1163

\bibitem[{{Sparre} {et~al.}(2011){Sparre}, {Sollerman}, {Fynbo}, {Malesani},
  {Goldoni}, {de Ugarte Postigo}, {Covino}, {D'Elia}, {Flores}, {Hammer},
  {Hjorth}, {Jakobsson}, {Kaper}, {Leloudas}, {Levan}, {Milvang-Jensen},
  {Schulze}, {Tagliaferri}, {Tanvir}, {Watson}, {Wiersema}, \&
  {Wijers}}]{2011ApJ...735L..24S}
{Sparre}, M., {et~al.} 2011, \apjl, 735, L24

\bibitem[{{Stanek} {et~al.}(2003){Stanek}, {Matheson}, {Garnavich}, {Martini},
  {Berlind}, {Caldwell}, {Challis}, {Brown}, {Schild}, {Krisciunas}, {Calkins},
  {Lee}, {Hathi}, {Jansen}, {Windhorst}, {Echevarria}, {Eisenstein}, {Pindor},
  {Olszewski}, {Harding}, {Holland}, \& {Bersier}}]{2003ApJ...591L..17S}
{Stanek}, K.~Z., {et~al.} 2003, \apjl, 591, L17

\bibitem[{{Tanvir} {et~al.}(2009){Tanvir}, {Fox}, {Levan}, {Berger},
  {Wiersema}, {Fynbo}, {Cucchiara}, {Kr{\"u}hler}, {Gehrels}, {Bloom},
  {Greiner}, {Evans}, {Rol}, {Olivares}, {Hjorth}, {Jakobsson}, {Farihi},
  {Willingale}, {Starling}, {Cenko}, {Perley}, {Maund}, {Duke}, {Wijers},
  {Adamson}, {Allan}, {Bremer}, {Burrows}, {Castro-Tirado}, {Cavanagh}, {de
  Ugarte Postigo}, {Dopita}, {Fatkhullin}, {Fruchter}, {Foley}, {Gorosabel},
  {Kennea}, {Kerr}, {Klose}, {Krimm}, {Komarova}, {Kulkarni}, {Moskvitin},
  {Mundell}, {Naylor}, {Page}, {Penprase}, {Perri}, {Podsiadlowski}, {Roth},
  {Rutledge}, {Sakamoto}, {Schady}, {Schmidt}, {Soderberg}, {Sollerman},
  {Stephens}, {Stratta}, {Ukwatta}, {Watson}, {Westra}, {Wold}, \&
  {Wolf}}]{2009Natur.461.1254T}
{Tanvir}, N.~R., {et~al.} 2009, \nat, 461, 1254

\bibitem[{{Tanvir} {et~al.}(2012){Tanvir}, {Levan}, {Fruchter}, {Fynbo},
  {Hjorth}, {Wiersema}, {Bremer}, {Rhoads}, {Jakobsson}, {O'Brien}, {Stanway},
  {Bersier}, {Natarajan}, {Greiner}, {Watson}, {Castro-Tirado}, {Wijers},
  {Starling}, {Misra}, {Graham}, \& {Kouveliotou}}]{2012ApJ...754...46T}
---. 2012, \apj, 754, 46

\bibitem[{{Th{\"o}ne} {et~al.}(2013){Th{\"o}ne}, {Fynbo}, {Goldoni}, {de
  Ugarte}, {Campana}, {Vergani}, {Covino}, {Kr{\"u}hler}, {Kaper}, {Tanvir},
  {Zafar}, {D'Elia}, {Gorosabel}, {Greiner}, {Groot}, {Hammer}, {Jakobsson},
  {Klose}, {Levan}, {Milvang-Jensen}, {Nicuesa}, {Palazzi}, {Piranomonte},
  {Tagliaferri}, {Watson}, {Wiersema}, \& {Wijers}}]{2013MNRAS.428.3590T}
{Th{\"o}ne}, C.~C., {et~al.} 2013, \mnras, 428, 3590

\bibitem[{{Th{\"o}ne} {et~al.}(2008){Th{\"o}ne}, {Wiersema}, {Ledoux},
  {Starling}, {de Ugarte Postigo}, {Levan}, {Fynbo}, {Curran}, {Gorosabel},
  {van der Horst}, {Llorente}, {Rol}, {Tanvir}, {Vreeswijk}, {Wijers}, \&
  {Kewley}}]{2008A&A...489...37T}
---. 2008, \aap, 489, 37

\bibitem[{{Tolstoy}(2011)}]{2011Sci...333..176T}
{Tolstoy}, E. 2011, Science, 333, 176

\bibitem[{{Vernet} {et~al.}(2011){Vernet}, {Dekker}, {D'Odorico}, {Kaper},
  {Kjaergaard}, {Hammer}, {Randich}, {Zerbi}, {Groot}, {Hjorth}, {Guinouard},
  {Navarro}, {Adolfse}, {Albers}, {Amans}, {Andersen}, {Andersen}, {Binetruy},
  {Bristow}, {Castillo}, {Chemla}, {Christensen}, {Conconi}, {Conzelmann},
  {Dam}, {de Caprio}, {de Ugarte Postigo}, {Delabre}, {di Marcantonio},
  {Downing}, {Elswijk}, {Finger}, {Fischer}, {Flores}, {Fran{\c c}ois},
  {Goldoni}, {Guglielmi}, {Haigron}, {Hanenburg}, {Hendriks}, {Horrobin},
  {Horville}, {Jessen}, {Kerber}, {Kern}, {Kiekebusch}, {Kleszcz}, {Klougart},
  {Kragt}, {Larsen}, {Lizon}, {Lucuix}, {Mainieri}, {Manuputy}, {Martayan},
  {Mason}, {Mazzoleni}, {Michaelsen}, {Modigliani}, {Moehler}, {M{\o}ller},
  {Norup S{\o}rensen}, {N{\o}rregaard}, {P{\'e}roux}, {Patat}, {Pena}, {Pragt},
  {Reinero}, {Rigal}, {Riva}, {Roelfsema}, {Royer}, {Sacco}, {Santin},
  {Schoenmaker}, {Spano}, {Sweers}, {Ter Horst}, {Tintori}, {Tromp}, {van
  Dael}, {van der Vliet}, {Venema}, {Vidali}, {Vinther}, {Vola}, {Winters},
  {Wistisen}, {Wulterkens}, \& {Zacchei}}]{2011A&A...536A.105V}
{Vernet}, J., {et~al.} 2011, \aap, 536, A105

\bibitem[{{Vreeswijk} {et~al.}(2004){Vreeswijk}, {Ellison}, {Ledoux}, {Wijers},
  {Fynbo}, {M{\o}ller}, {Henden}, {Hjorth}, {Masi}, {Rol}, {Jensen}, {Tanvir},
  {Levan}, {Castro Cer{\'o}n}, {Gorosabel}, {Castro-Tirado}, {Fruchter},
  {Kouveliotou}, {Burud}, {Rhoads}, {Masetti}, {Palazzi}, {Pian}, {Pedersen},
  {Kaper}, {Gilmore}, {Kilmartin}, {Buckle}, {Seigar}, {Hartmann}, {Lindsay},
  \& {van den Heuvel}}]{2004A&A...419..927V}
{Vreeswijk}, P.~M., {et~al.} 2004, \aap, 419, 927

\bibitem[{{Vreeswijk} {et~al.}(2007){Vreeswijk}, {Ledoux}, {Smette}, {Ellison},
  {Jaunsen}, {Andersen}, {Fruchter}, {Fynbo}, {Hjorth}, {Kaufer}, {M{\o}ller},
  {Petitjean}, {Savaglio}, \& {Wijers}}]{vreeswijk2007}
---. 2007, \aap, 468, 83

\bibitem[{{Watson}(2011)}]{Watson11}
{Watson}, D. 2011, \aap, 533, A16

\bibitem[{{Watson} {et~al.}(2006){Watson}, {Fynbo}, {Ledoux}, {Vreeswijk},
  {Hjorth}, {Smette}, {Andersen}, {Aoki}, {Augusteijn}, {Beardmore}, {Bersier},
  {Castro Cer{\'o}n}, {D'Avanzo}, {Diaz-Fraile}, {Gorosabel}, {Hirst},
  {Jakobsson}, {Jensen}, {Kawai}, {Kosugi}, {Laursen}, {Levan}, {Masegosa},
  {N{\"a}r{\"a}nen}, {Page}, {Pedersen}, {Pozanenko}, {Reeves}, {Rumyantsev},
  {Shahbaz}, {Sharapov}, {Sollerman}, {Starling}, {Tanvir}, {Torstensson}, \&
  {Wiersema}}]{2006ApJ...652.1011W}
{Watson}, D., {et~al.} 2006, \apj, 652, 1011

\bibitem[{{Wiersema} {et~al.}(2011){Wiersema}, {Flores}, {D'Elia}, {Goldoni},
  {Malesani}, {Vergani}, {de Ugarte Postigo}, {Levan}, {Milvang-Jensen}, \&
  {Fynbo}}]{2011GCN..12431...1W}
{Wiersema}, K., {et~al.} 2011, GRB Coordinates Network, 12431, 1

\bibitem[{{Willingale} {et~al.}(2013){Willingale}, {Starling}, {Beardmore},
  {Tanvir}, \& {O'Brien}}]{2013MNRAS.431..394W}
{Willingale}, R., {Starling}, R.~L.~C., {Beardmore}, A.~P., {Tanvir}, N.~R., \&
  {O'Brien}, P.~T. 2013, \mnras, 431, 394

\bibitem[{{Wolfe} {et~al.}(2005){Wolfe}, {Gawiser}, \&
  {Prochaska}}]{2005ARA&A..43..861W}
{Wolfe}, A.~M., {Gawiser}, E., \& {Prochaska}, J.~X. 2005, \araa, 43, 861

\bibitem[{{Wolfe} {et~al.}(2008){Wolfe}, {Prochaska}, {Jorgenson}, \&
  {Rafelski}}]{2008ApJ...681..881W}
{Wolfe}, A.~M., {Prochaska}, J.~X., {Jorgenson}, R.~A., \& {Rafelski}, M. 2008,
  \apj, 681, 881

\bibitem[{{Xu} {et~al.}(2013){Xu}, {de Ugarte Postigo}, {Leloudas},
  {Kr{\"u}hler}, {Cano}, {Hjorth}, {Malesani}, {Fynbo}, {Th{\"o}ne},
  {S{\'a}nchez-Ram{\'{\i}}rez}, {Schulze}, {Jakobsson}, {Kaper}, {Sollerman},
  {Watson}, {Cabrera-Lavers}, {Cao}, {Covino}, {Flores}, {Geier}, {Gorosabel},
  {Hu}, {Milvang-Jensen}, {Sparre}, {Xin}, {Zhang}, {Zheng}, \&
  {Zou}}]{2013arXiv1305.6832X}
{Xu}, D., {et~al.} 2013, \apj, 776, 98

\bibitem[{{Xu} {et~al.}(2011){Xu}, {Malesani}, {Buchhave}, {Schulze}, \&
  {Jakobsson}}]{2011GCN..12427...1X}
{Xu}, D., {Malesani}, D., {Buchhave}, L.~A., {Schulze}, S., \& {Jakobsson}, P.
  2011, GRB Coordinates Network, 12427, 1

\bibitem[{{Zafar} \& {Watson}(2013)}]{2013arXiv1303.1141Z}
{Zafar}, T., \& {Watson}, D. 2013, \aap, 560, A26

\bibitem[{{Zafar} {et~al.}(2011){Zafar}, {Watson}, {Fynbo}, {Malesani},
  {Jakobsson}, \& {de Ugarte Postigo}}]{2011A&A...532A.143Z}
{Zafar}, T., {Watson}, D., {Fynbo}, J.~P.~U., {Malesani}, D., {Jakobsson}, P.,
  \& {de Ugarte Postigo}, A. 2011, \aap, 532, A143

\end{thebibliography}

\clearpage

\appendix

\section{The issue of saturation for the S II 1253-transition}\label{S_saturation}

In Section~\ref{section:chempos} it is mentioned that the \ion{S}{2} $\lambda$1253 transition is mildly saturated. To clearly address the issue of saturation for this transition we developed a multi-component line fitting code, where the Monte Carlo Markov Chain (MCMC) package, \textit{Emcee} \citep{2013PASP..125..306F}, is used as fitting method. Such MCMC-algorithms are ideal for cases with degenerate models. A fit is performed for \ion{Ni}{2} $\lambda$1370 and \ion{S}{2} $\lambda$1253. In the MCMC run we use a flat prior with $8<b($km s$^{-1})<40$ and $12<\log N /$cm$^{-2}<19$ for both S and Ni. Again, we use two absorption components for each transition, and the $b$- and $z$-values are linked between \ion{Ni}{2} $\lambda$1370 and \ion{S}{2} $\lambda$1253. We sample 96000 points from the likelihood function after a burn-in of 32000 points (to ensure convergence of the chain).

In Fig. \ref{mcmccontours} the 1$\sigma$, 2$\sigma$ and 3$\sigma$ contours are shown for the $b$- and $N$-values, as well as contours of constant equivalent width. For \ion{Ni}{2} $\lambda$1370 the column density is well-constrained for both components\footnote{Note that here the contribution from fine structure lines is not included in the column density of nickel (unlike in Table~\ref{table:metallicity}, where it is included).}. For \ion{S}{2} $\lambda$1253 the blue component has a well-constrained column density, whereas the red component is saturated (the column density can vary by four orders of magnitude within the $2\sigma$ confidence interval).

For \ion{Ni}{2} $\lambda$1370  the difference in $\log N$ for the two components is relatively small (for the blue component $\log N / $cm$^{-2}$ is only 0.1 dex lower than for the red component). Assuming the same difference in $\log N / $cm$^{-2}$ for the two components of \ion{S}{2} $\lambda$1253, we derive $\log N / $cm$^{-2} = 15.79 \pm 0.08 $, which is clearly consistent with the value reported in Table~\ref{table:metallicity}. Thus, the column density of \ion{S}{2} can be reliably determined even though one of the components is saturated.

Finally, we note that the blue component has a larger $b$-value than the red component in the model in the present section. This is contradicting the results from Section~\ref{section:chempos}, where the red component has a larger $b$-value than the blue component. The explanation is that more absorption lines are included in the model in Section~\ref{section:chempos}. If we remove all other absorption components than \ion{S}{2} and \ion{Ni}{2} from the model in Section~\ref{section:chempos} we get $b=22.8\pm4.0$ km s$^{-1}$ for the blue component and $b=15.6\pm 4.3$ km s$^{-1}$ for the red component, which is consistent with the $b$-values from the present section. In all cases (i.e. in the MCMC model, and in the two VPFIT models) we obtain column densities which are consistent with each other, so this is not an important caveat.

\subsection{The effect a third absorption component}

An assumption in the above derivation of the sulfur column density is that there are only two absorption components. We will now discuss how a third hidden component would affect the derived column density of sulfur. The red component of \ion{S}{2} $\lambda$1253 is mildly saturated, and it would therefore be possible to add a narrow line with the same redshift as the red component without affecting the way the spectrum looks. We could for example add an extra component with $b=10$ km s$^{-1}$ and $\log N/$cm$^{-2}=16.3$ without affecting the spectrum, if the component has a redshift coinciding with the red component.

\begin{figure}
\centering
\includegraphics[width=\linewidth]{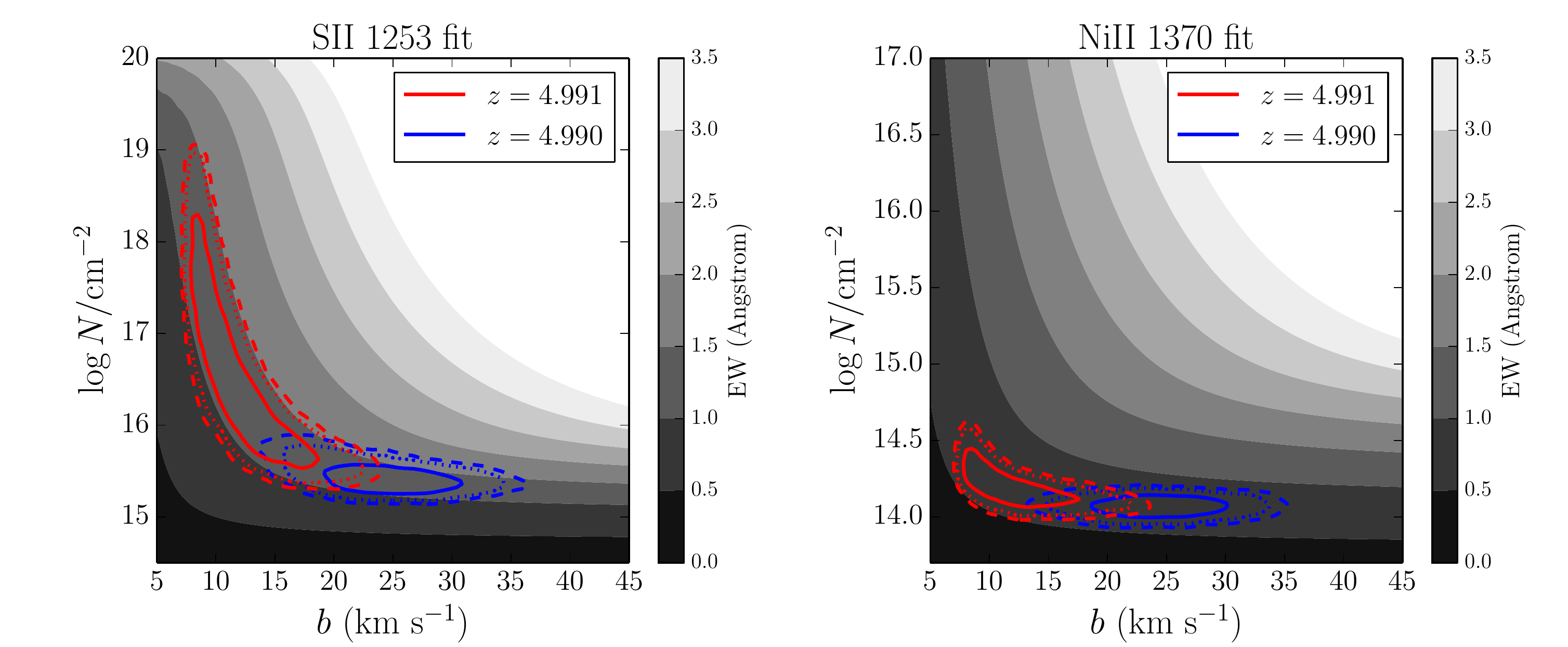}
\caption{1$\sigma$, 2$\sigma$ and 3$\sigma$ confidence levels (solid, dotted and dashed contours, respectively) for the Doppler-parameters and the column densities for \ion{S}{2} and \ion{Ni}{2}. The grey contours show regions of constant equivalent width.}
\label{mcmccontours}
\end{figure}

The blue component is not saturated, so here it is not possible to add such a component without heavily modifying the spectrum. The situation is therefore the same as in Fig.~\ref{mcmccontours}; the blue component is reliably determined, and the red component is uncertain because it is saturated, so we can only derive the total column density under the assumption that the ratio between the column density of the blue and the red component is the same as for nickel. Our conclusion is that the probability that large column densities are present in a hidden saturated component is low.

\FloatBarrier

\clearpage
\section{The X-shooter spectrum}\label{xsh-appendix}

\begin{figure}
\centering
\includegraphics[width=\linewidth]{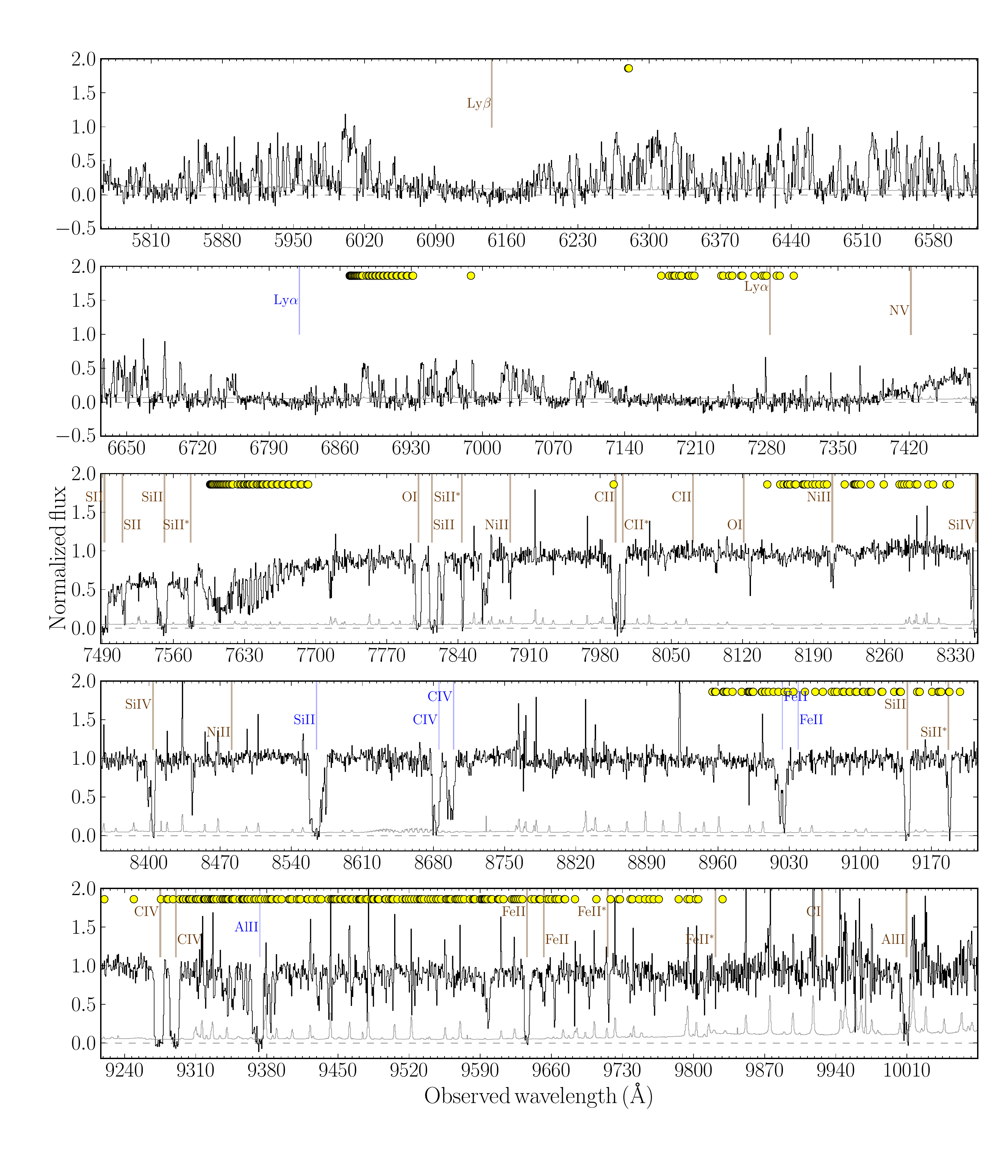}
\caption{The VIS spectrum. The yellow circles mark telluric features, the brown marks show absorption lines from the GRB host galaxy, and the blue marks show absorption features from the intervening system at $z=4.6$. The spectrum is shown in black, and the error spectrum is shown in gray.}
\label{specvis}
\end{figure}

\begin{figure}
\centering
\includegraphics[width=\linewidth]{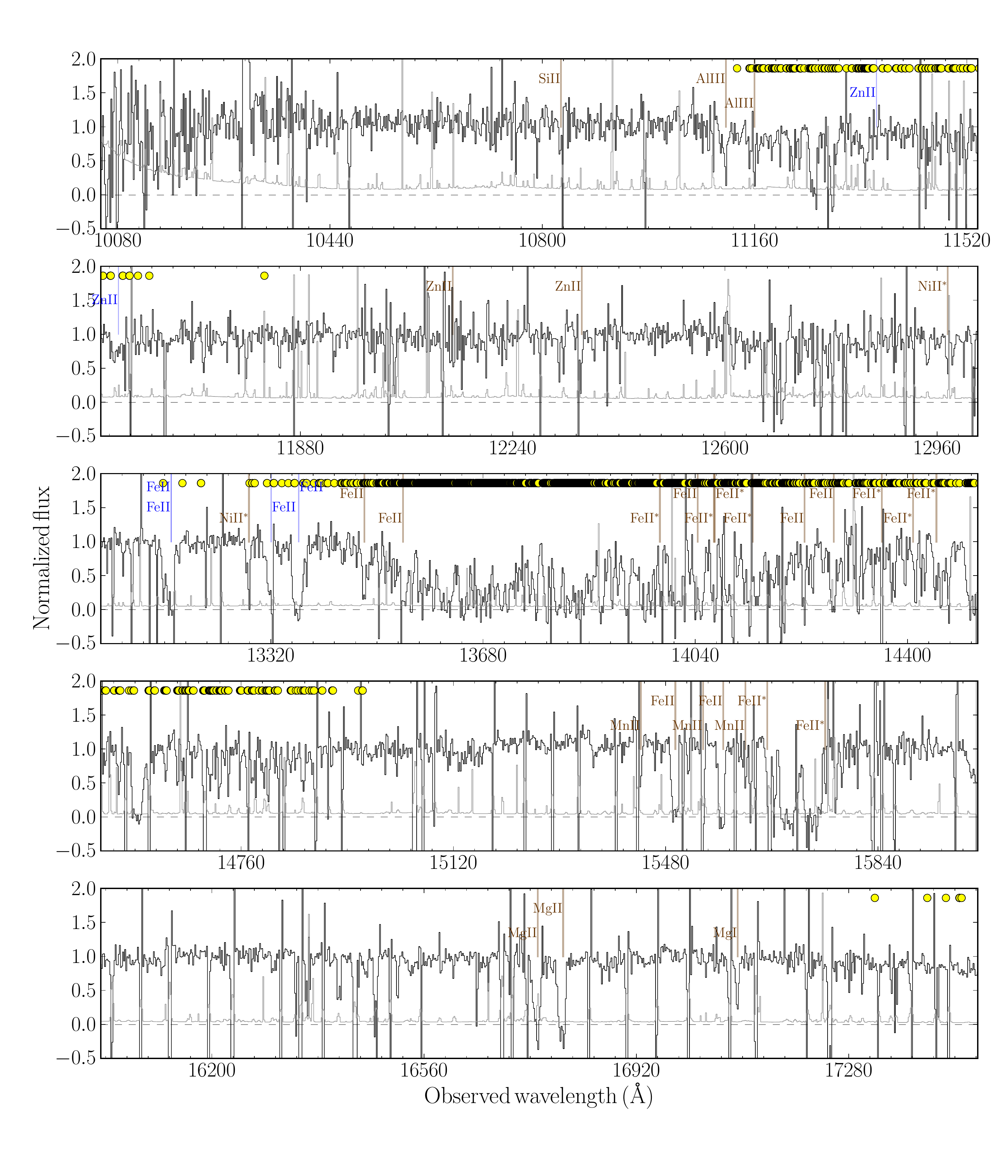}
\caption{Same as Fig.~\ref{specvis}, but for the NIR spectrum.}
\label{specnir}
\end{figure}

Fig. \ref{specvis} and \ref{specnir} show absorption lines in the normalized X-shooter spectrum for the VIS and NIR arms. 
\FloatBarrier

\end{document}